\documentclass[12pt]{article}

\usepackage{array} 
\usepackage{epsfig}
\usepackage{amssymb}
\usepackage{amsmath}                          
\usepackage{graphics,graphpap}

\renewcommand{\thefootnote}{\fnsymbol{footnote}}

\newcommand{\rmega}{\omega}
\newcommand{\rhi}{\phi}
\newcommand{\rar}{\bar}
\newcommand{\relta}{\delta}
\newcommand{\rline}{\hline}

\newcommand{\bra}{\langle}
\newcommand{\ket}{\rangle}

\newcommand{\wt}{\widetilde}

\def\s#1{\setbox0=\hbox{$#1$}%
  \rlap{\ifdim\wd0>.7em\kern.22\wd0\else\kern.1\wd0\fi /}#1}
\newcommand{\matel}[3]{\langle #1|#2|#3\rangle}
\newcommand{\al}{\alpha}
\newcommand{\ga}{\gamma}

\def\pslash{\rlap/{\mkern-1mu p}}
\def\zslash{\rlap/{\mkern-1mu z}}
\makeatletter
\def\slash#1{{\mathpalette\c@ncel{#1}}} 
\makeatother
                                                                               
\begin{document}

\begin{titlepage}
\begin{flushright}
\begin{tabular}{l}
IPPP/04/74\\
DCPT/04/48\\
TPI-MINN-04/39\\
\end{tabular}
\end{flushright}
\vskip1.5cm
\begin{center}
   {\Large \bf \boldmath 
     $B_{d,s}\to\rho,\omega,
K^*,\phi$ Decay  Form Factors\\[5pt] from
    \text{Li}ght-Cone Sum Rules Revisited}
    \vskip1.3cm {\sc
Patricia Ball\footnote{Patricia.Ball@durham.ac.uk}$^{,1}$ and 
    Roman Zwicky\footnote{zwicky@physics.umn.edu}}$^{,2}$
  \vskip0.5cm
        $^1$ IPPP, Department of Physics, 
University of Durham, Durham DH1 3LE, UK \\
\vskip0.4cm 
$^2$ William I. Fine Theoretical Physics Institute, \\ University of
Minnesota, Minneapolis, MN 55455, USA  
\vskip2cm


\vskip3cm

{\large\bf Abstract:\\[10pt]} \parbox[t]{\textwidth}{ 
We present an improved calculation of $B\to$ light vector
form factors from light-cone sum rules, including one-loop radiative
corrections to twist-2 and twist-3 contributions, and leading order
twist-4 corrections. The total theoretical uncertainty of our
results at zero momentum transfer is typically 10\% and can be improved,
at least in part, by
reducing the uncertainty of hadronic input parameters. 
We present our results in a way which details the
dependence of the form factors on these parameters and facilitates the
incorporation of future updates of their values from
e.g.\ lattice calculations. We also give simple and
easy-to-inplement parametrizations of the $q^2$-dependence of the
form factors which are valid in the full kinematical regime of $q^2$.
}

\vfill

{\em submitted to Physical Review D}
\end{center}
\end{titlepage}

\setcounter{footnote}{0}
\renewcommand{\thefootnote}{\arabic{footnote}}

\newpage

\section{Introduction}\label{sec:intro}

This paper aims to give
a new and more precise determination of the
decay form factors of $B_{d,u,s}$ mesons into light vector mesons, i.e.\
$\rho$, $\omega$, $K^*$ and $\phi$; it is a continuation of of our
previous study of $B$ decays into pseudoscalar mesons \cite{BZ}.
The calculation uses the method of QCD sum
rules on the light-cone, which in the past has been rather
successfully applied to various problems in heavy-meson physics,
cf.~Refs.~\cite{protz};\footnote{See also
  Ref.~\cite{LCSRs:reviews} for reviews.}
an outline of the method will be given below. Our
calculation improves on our previous paper \cite{BB98} by
\begin{itemize}
\item including $B\to\omega$ form factors;
\item including radiative corrections to 2-particle twist-3 contributions to
  one-loop accuracy;
\item a new parametrization of the dominant hadronic contributions
  (twist-2 distribution amplitudes);
\item detailing the dependence of form factors on distribution
  amplitudes;
\item a new parametrization of the dependence of the form factors on
  momentum transfer;
\item a careful analysis of the theoretical uncertainties.
\end{itemize}
Like in Ref.~\cite{BZ},
the motivation for this study is twofold and relates to the overall
aim of $B$ physics to provide precision determinations of quark flavor
mixing parameters in the Standard Model. Quark flavor mixing is
governed by the unitary CKM matrix which depends on four parameters: three
angles and one phase. The constraints from unitarity can be visualized
by the so-called unitarity triangles (UT); the one that is relevant
for $B$ physics is under intense experimental study. 
The over-determination of the sides
and angles of this triangle from a multitude of processes will answer
the question whether there is new physics in flavor-changing
processes and where it manifests itself. One of the sides of the UT is
given by the ratio of CKM matrix elements $|V_{ub}/V_{cb}|$.
$|V_{cb}|$ is known to about 2\% accuracy from both inclusive and
exclusive $b\to c\ell \nu$ transitions \cite{CKMWS}, whereas the
present error
on $|V_{ub}|$ is much larger and around $15\%$. Its reduction requires
an improvement of experimental statistics, which is under way at the $B$
factories BaBar and Belle, but also and in particular an improvement
of the theoretical prediction for associated semileptonic spectra and
decay rates. This is one motivation for our study of the
$B\to\rho$ semileptonic decay form factors $A_1$, $A_2$, $V$, 
which, in conjunction with
alternative calculations, hopefully from lattice,
will help to reduce the uncertainty from exclusive semileptonic 
determinations of $|V_{ub}|$. Secondly, form factors of general
$B\to\,$light meson transitions are also needed as ingredients in the
analysis of nonleptonic two-body $B$ decays, e.g.\ $B\to \rho\pi$, 
in the framework of QCD factorization
\cite{QCDfac}, again with the objective to extract CKM parameters. 
 One issue calling for particular attention in this context
is the effect of SU(3) breaking, which enters
both the form factors and the $K^*$ and $\phi$ meson distribution
amplitudes figuring in the factorization analysis. We would like to
point out that the implementation of SU(3) breaking in the
light-cone sum rules approach to form factors is {\em precisely} the
same as in QCD factorization and is encoded in the difference between
$\rho$, $\omega$, $K^*$ and $\phi$ distribution amplitudes, so that the use of
form factors calculated from light-cone sum rules together with the
corresponding meson distribution amplitudes in factorization
formulas allows a unified
and controlled approach to the assessment of SU(3) breaking effects in
nonleptonic $B$ decays.

As we shall detail below, 
QCD sum rules on the light-cone allow the calculation of form factors
in a kinematic regime where the final state meson has large energy in
the rest-system of the decaying $B$,
$E\gg \Lambda_{\rm QCD}$. 
The physics underlying $B$ decays into light mesons at large momentum
transfer can be understood qualitatively in the framework of hard
exclusive QCD processes, pioneered by Brodsky and Lepage et al.\ 
\cite{pQCD}. The hard scale in $B$ decays is $m_b$ and one can show
that to leading order in $1/m_b$ the decay is described by two
different parton configurations: one 
where all quarks have large momenta and the momentum transfer
happens via the exchange of a hard gluon, the so-called hard-gluon
exchange, and a second one where one quark is soft and does
interact with the other partons only via soft-gluon exchange, the
so-called soft or Feynman-mechanism. The consistent treatment of both
effects in a framework based on factorization, i.e.\ the clean separation of
perturbatively calculable hard contributions from nonperturbative
``wave functions'', is highly nontrivial and has spurred the
development of SCET, an effective field theory which aims to separate
the two relevant large mass scales $m_b$ and $\sqrt{m_b\Lambda_{\rm
    QCD}}$ in a systematic way \cite{SCET}. 
In this approach form factors can indeed be split into
a calculable factorizable part which roughly corresponds to the hard-gluon
exchange contributions, and a nonfactorizable one, which includes the soft
contributions and cannot be calculated within the SCET
framework \cite{fuck,hill}. 
Predictions obtained in this approach then typically aim to eliminate the soft
part and take the form of relations
between two or more form factors whose difference is expressed in terms
of factorizable contributions. 

The above discussion highlights the need for a calculational
method that allows numerical predictions while treating both 
hard and soft contributions on the same footing. It is precisely QCD
sum rules on the light-cone (LCSRs) that accomplish this task.
LCSRs can be viewed as an extension of the original 
method of QCD sum rules devised by
Shifman, Vainshtein and Zakharov (SVZ) \cite{SVZ}, which was
designed to determine properties of
ground-state hadrons at
zero or low momentum transfer, to the regime of large
momentum transfer. QCD sum rules combine the
concepts of operator product expansion, dispersive representations of
correlation functions and quark-hadron duality in an ingenious way
that allows the calculation of the properties of non-excited 
hadron-states with a very reasonable theoretical uncertainty. 
In the context of weak-decay form factors, the basic quantity is the 
correlation function of the weak current and a 
current with
the quantum numbers of the $B$ meson, evaluated between the vacuum and
a light meson.
For large (negative) virtualities of these currents, the
correlation function is, in coordinate-space, dominated by distances
close to the light-cone and can be discussed in the framework of
light-cone expansion. In contrast to the short-distance expansion
employed by conventional QCD sum rules \`a la SVZ where
nonperturbative effects are encoded in vacuum expectation values
of local operators with
vacuum quantum numbers, the condensates, LCSRs
rely on the factorization of the underlying correlation function into
genuinely nonperturbative and universal hadron distribution amplitudes (DAs)
$\phi$ which are convoluted with process-dependent amplitudes $T$.
The latter are the analogues of Wilson-coefficients in the
short-distance expansion and can be
calculated in perturbation theory. The light-cone expansion then reads, 
schematically:
\begin{equation}\label{eq:schemat}
\mbox{correlation function~}\sim \sum_n T^{(n)}\otimes \phi^{(n)}.
\end{equation}
The sum runs over contributions with increasing twist, labelled by
$n$, which are suppressed by
increasing powers of, roughly speaking, the virtualities of the
involved currents.
The same correlation function can, on the other hand, be written as a
dispersion-relation, in the virtuality of the current coupling to the
$B$ meson. Equating dispersion-representation and the
light-cone expansion, and separating the $B$ meson contribution from
that of higher one- and multi-particle states using quark-hadron
duality, one obtains a relation
for the form factor describing the decay $B\to\,$light meson.

A crucial question is the accuracy of light-cone sum
rules. Like with most other methods, there are 
uncertainties induced by external parameters like quark masses and
hadronic parameters and intrinsic uncertainties 
induced by the approximations inherent in
the method. As we shall discuss in Sec.~4, the total theoretical
uncertainty for the form factors at $q^2=0$ is presently around 10\%,
including a 7\% irreducible systematic  uncertainty.

Our paper is organized as follows: in Sec.~2 we define all relevant
quantities, in particular the meson distribution
amplitudes. In Sec.~3 we outline the calculation. 
In Sec.~4 we derive the sum rules and present  numerical
results which are summarized in Tabs.~\ref{tab:basic} and
\ref{tab:fits}. Section~5 contains a summary and conclusions. Detailed
expressions for distribution amplitudes, a break-down of the
light-cone sum rule results into different contributions and explicit
formulas for the contributions of 3-particle states 
are given in the appendices.

\section{Definitions}\label{sec:2}

$B\to V$ transitions, where $V$ stands for the vector mesons $\rho$,
$\omega$, $K^*$ and $\phi$, can manifest themselves as 
semileptonic decays $B\to V \ell \bar \nu_\ell$ or rare
flavor-changing neutral current (FCNC) penguin-induced decays $B\to V\gamma$ and
$B\to V \ell^+ \ell^-$. All these decays are described by a
total of seven independent form factors which usually are defined as 
($q=p_B-p$)
\begin{eqnarray}
\lefteqn{c_V
\langle V(p) | \bar q\gamma_\mu(1-\gamma_5) b | B(p_B)\rangle  =  
-i e^*_\mu (m_B+m_V)
A_1(q^2) + i (p_B+p)_\mu (e^* q)\,
\frac{A_2(q^2)}{m_B+m_V}}\hspace*{2.8cm}\nonumber\\
&& {}+  i
q_\mu (e^* q) \,\frac{2m_V}{q^2}\,
\left(A_3(q^2)-A_0(q^2)\right) +
\epsilon_{\mu\nu\rho\sigma}\epsilon^{*\nu} p_B^\rho p^\sigma\,
\frac{2V(q^2)}{m_B+m_V}\hspace*{0.5cm}\label{eq:SLFF}\\
{\rm with\ }A_3(q^2) & = & \frac{m_B+m_V}{2m_V}\, A_1(q^2) -
\frac{m_B-m_V}{2m_V}\, A_2(q^2)\mbox{~~and~~} 
A_0(0) =  A_3(0);\label{eq:A30}\\[-1cm]\nonumber
\end{eqnarray}
\begin{eqnarray}
\lefteqn{c_V\langle V(p) | \bar q \sigma_{\mu\nu} q^\nu (1+\gamma_5) b |
B(p_B)\rangle = i\epsilon_{\mu\nu\rho\sigma} \epsilon^{*\nu}
p_B^\rho p^\sigma \, 2 T_1(q^2)}\nonumber\\
& & {} + T_2(q^2) \left\{ e^*_\mu
  (m_B^2-m_{V}^2) - (e^* q) \,(p_B+p)_\mu \right\} + T_3(q^2) 
(e^* q) \left\{ q_\mu - \frac{q^2}{m_B^2-m_{V}^2}\, (p_B+p)_\mu
\right\}\hspace*{0.4cm}\label{eq:pengFF}\\
\lefteqn{{\rm with\ } T_1(0) = T_2(0). \label{eq:T1T2}}
\end{eqnarray}
$A_0$ is also the form factor of the pseudoscalar current:
\begin{equation}\label{eq:A0}
c_V\langle V |\partial_\mu A^\mu | B\rangle = c_V(m_b+m_q)
\langle V |\bar q i\gamma_5 b | B\rangle = 2 m_V
(e^* q) A_0(q^2).
\end{equation}
$\bar q$ in the above formulas stands for $\bar u$, $\bar d$ and $\bar s$
in (\ref{eq:SLFF}) and (\ref{eq:A0}) 
and $\bar d$, $\bar s$ in (\ref{eq:pengFF}); the
actual assignment of different decay channels to underlying $b\to
q$ transitions is made explicit in Tab.~\ref{tab:0}. In our
calculation, we assume isospin symmetry throughout, which implies 
that there are five different
sets of form factors: $B_q\to\rho$, $B_q\to\omega$, $B_q\to K^*$, 
$B_s\to  K^*$ and $B_s\to \phi$ (with $q=u,d$).
The factor $c_V$ accounts for the flavor content of 
particles: $c_V=\sqrt{2}$ for $\rho^0$,
$\omega$ and $c_V=1$ otherwise.\footnote{To be
  precise, $c_V$ is $\sqrt{2}$ for $\rho^0$ in
  $b\to u$ and for $\omega$, and  $-\sqrt{2}$ for $\rho^0$ in
  $b\to d$, with the flavor wave functions $\rho^0\sim (\bar u u -
  \bar d d)/\sqrt{2}$ and $\omega \sim (\bar u u +
  \bar d d)/\sqrt{2}$. We assume that $\phi$ is a pure $s\bar s$ state.}

The currents in (\ref{eq:SLFF}) and (\ref{eq:pengFF}) contain
both vector and axialvector components. $V$ and $T_1$
correspond to the vector components of the currents, and, as the $B$
meson is a pseudoscalar, to 
the axialvector components of the matrix elements. 
$A_{1,2}$ clearly correspond to the axialvector component of the $V-A$
current; the term in $A_3-A_0$ arises as the contraction of
(\ref{eq:SLFF}) with $-iq^\mu$ must agree with (\ref{eq:A0}). As for
the penguin current,  $T_{2,3}$  correspond to the
axialvector components of the current; there is no analogon to $A_0$,
as the  current vanishes upon contraction with $q^\mu$. 
As we shall see in Sec.~\ref{sec:4}, for analysing the dependence of
each form factor on $q^2$, it is best to choose $A_{0,1,2}$ as
independent form factors for the $A$ current, and define $A_3$ by
(\ref{eq:A30}), but for the penguin
current it will turn out  more appropriate to choose a different set
of independent form factors:  $T_1$,
$T_2$ and $\widetilde{T}_3$ with
\begin{eqnarray}
\lefteqn{c_V\langle V(p) | \bar q \sigma_{\mu\nu} q^\nu (1+\gamma_5) b |
B(p_B)\rangle = i\epsilon_{\mu\nu\rho\sigma} \epsilon^{*\nu}
p_B^\rho p^\sigma \, 2 T_1(q^2)}\nonumber\\
& & {} + e^*_\mu (m_B^2-m_{V}^2)  T_2(q^2) -(p_B+p)_\mu (e^* q)
\widetilde{T}_3(q^2) + q_\mu (e^* q) T_3(q^2)\label{eq:pengFF2}
\end{eqnarray}
and $T_3$ defined as
\begin{equation}
 T_3(q^2) =\frac{m_B^2-m_V^2}{q^2}\left(
\widetilde{T}_3(q^2) -  T_2(q^2)\right).\label{eq:T3tilde}
\end{equation}

As the actual calculation is done using an off-shell momentum $p_B$
with $p_B^2\neq m_B^2$, it is
crucial to avoid any ambiguity in the interpretation of scalar
products like $2pq=p_B^2-q^2-p^2\neq m_B^2-q^2-m_V^2$ that occur at 
intermediate steps of
the calculation. This is particularly relevant for the penguin form
factors which are defined in terms of a matrix element over the tensor
current which is
contracted with the physical momentum $q^\nu$. The problem can be
avoided by  extracting
 $T_i$ and $\widetilde{T}_3$ from sum
rules for a matrix element with no contractions:
\begin{eqnarray}
\langle V(p)|\bar q \sigma_{\mu\nu}\gamma_5 b | B(p_B)\rangle & = &
A(q^2) \left\{e^*_\mu (p_B+p)_\nu - (p_B+p)_\mu e^*_\nu\right\} - B(q^2)
\left\{e^*_\mu q_\nu - q_\mu e^*_\nu\right\}\nonumber\\
& & {} -  2C(q^2) \,\frac{e^*
  q}{m_B^2-m_{V}^2} \,\left\{ p_\mu q_\nu - q_\mu
  p_\nu \right\}.\label{6}
\end{eqnarray}
$A$, $B$ and $C$ are related to $T_i$ and $\widetilde{T}_3$ 
defined in (\ref{eq:pengFF}) and (\ref{eq:pengFF2}) as
\begin{equation}\label{eq:eq}
T_1 = A,\quad T_2 = A-\frac{q^2}{m_B^2-m_{V}^2}\, B,\quad
T_3 = B+C,\quad \widetilde{T}_3 = A + \frac{q^2}{m_B^2-m_{V}^2}\, C,
\end{equation}
which implies 
\begin{equation}
T_1(0)=T_2(0)=\widetilde{T}_3(0).
\end{equation}

\begin{table}[tb]
\addtolength{\arraycolsep}{3pt}
\renewcommand{\arraystretch}{1.2}
$$
\begin{array}{|l|cccccccc|}
\hline
 & \rho^+ & \rho^0,\omega & \rho^- & K^{*+} & K^{*0}(d\bar s) & K^{*-} &
 \bar{K}^{*0}(s\bar d) & \phi\\\hline
B_u^- & - & b\to u & b\to d & - & - & b\to s & - & -\\
\bar{B}_d & b\to u & b\to d & - & - & - & - & b\to s & -\\
\bar{B}_s & - & - & - & b\to u & b\to d & - & - & b\to s\\\hline 
\end{array}
$$
\addtolength{\arraycolsep}{-3pt}
\renewcommand{\arraystretch}{1}
\caption[]{Allowed decay channels in terms of underlying quark 
  transitions. We assume isospin-symmetry and hence have five different
  sets of form factors: $B_q\to\rho$, $B_q\to\omega$, $B_q\to K^*$, 
$B_s\to  K^*$ and $B_s\to \phi$ (with $q=u,d$).}\label{tab:0}
\end{table}

Relevant for semileptonic decays are, in the limit of vanishing lepton
mass, the form factors $A_{1,2}$ and $V$ with $q^2$, the invariant
mass of the lepton-pair, in the range $0\leq q^2\leq
(m_B-m_V)^2$. $B\to V\gamma$ depends on $T_1(0)$, whereas $B\to V
\ell^+\ell^-$ depends on all seven form factors (see Ref.~\cite{Hiller} for
an explicit formula). The motivation for studying $B\to\rho \ell \bar
\nu_\ell$ and $B\to\omega \ell \bar
\nu_\ell$ is to extract information on the CKM matrix element
$|V_{ub}|$, whereas the FCNC transitions $B\to(K^*,\rho,\omega)\gamma$
and $B\to(K^*,\rho,\omega)\ell^+\ell^-$ serve to constrain new
physics or, in the absence thereof, the ratio
$|V_{ts}/V_{td}|$ \cite{parkho}, which would complement the
  determination of $|V_{ts}/V_{td}|$ from $B$-mixing. 

In the LCSR approach the form factors are extracted from the
correlation function of the relevant weak current $J_W$, i.e.\ either
the semileptonic $V-A$ current or the penguin current of
(\ref{eq:T3tilde}), and an interpolating
field for the $B$ meson, in the presence of the vector meson:
\begin{equation}\label{eq:corr}
\Gamma(q^2,p_B^2) = 
i\int d^4x e^{iqx} \langle V(p) | T J_W(x) j_b^\dagger(0) | 0 \rangle,
\end{equation}
with $j_b = m_b \bar q' i \gamma_5 b$, $q'\in\{u,d,s\}$.
For virtualities
\begin{equation}\label{eq:virt}
m_b^2-p_B^2 \geq O(\Lambda_{\rm QCD}m_b), \qquad
m_b^2-q^2 \geq O(\Lambda_{\rm QCD}m_b),
\end{equation}
the correlation function \eqref{eq:corr}
is dominated by light-like distances
and therefore accessible to an expansion around the light-cone.
The above conditions can be understood by demanding that the exponential
factor in \eqref{eq:corr} vary only slowly. The light-cone expansion
is performed by integrating out the transverse and ``minus'' degrees
of freedom and leaving only the longitudinal momenta of the partons as
relevant degrees of freedom. The integration over transverse momenta
is done up to a cutoff, $\mu_{\rm IR}$, all momenta below which are
included in a so-called hadron distribution amplitude (DA) $\phi$, whereas
larger transverse momenta are calculated in perturbation theory. The
correlation function is hence decomposed, or factorized, into
perturbative contributions $T$ and nonperturbative contributions
$\phi$, which both depend on the longitudinal parton momenta and the
factorization scale $\mu_{\rm IR}$. If the vector meson is an effective
quark-antiquark bound state, as is the case to leading order in the
light-cone expansion, one can write the corresponding longitudinal
momenta as $up$ and $(1-u)p$, where $p$ is the momentum of the meson
and $u$ a number between 0 and 1. The schematic relation
\eqref{eq:schemat} can then be written
in more explicit form as
\begin{equation}
\label{eq:lcexp}
\Gamma(q^2,p_B^2) = \sum_n \int_0^1 du \,
T^{(n)}(u,q^2,p_B^2,\mu_{\text{IR}}) \phi^{(n)}(u,\mu_{\text{IR}}).
\end{equation}
As $\Gamma$ itself is independent of the arbitrary scale
$\mu_{\text{IR}}$, the scale-dependence of $T^{(n)}$ and $\phi^{(n)}$
must cancel
each other.\footnote{If there is more than one contribution of a
  given twist, they will mix under a change of the factorization scale
$\mu_{\rm IR}$ and it is only in the sum of all such contributions
  that the residual $\mu_{\rm IR}$ dependence cancels.}
If $\phi^{(n)}$ describes the meson in a 2-parton state, it is
called a 2-particle DA, if it describes
a 3-parton,
i.e.\ quark-antiquark-gluon state, it is called 3-particle DA. In
the latter case the integration over $u$ gets replaced by an
integration over two independent momentum fractions, say $\alpha_1$
and $\alpha_2$.
Eq.~(\ref{eq:lcexp}) is called a ``collinear'' factorization formula,
as the momenta of the partons in the meson are collinear with its momentum.
Any such factorisation formula requires verification
by explicit calculation; we will come back to
that issue in the next section.
                                                                               
Let us now define the distribution amplitudes to be used in this
paper. All definitions and  formulas are well-known
 and can be found in Ref.~\cite{wavefunctions}.
In general, the distribution amplitudes we are
interested in are related
to nonlocal matrix elements of type\footnote{The currents to use for
  $\rho^0$ and $\omega$ are $(\bar u\Gamma u \mp \bar d\Gamma
  d)/\sqrt{2}$, respectively.}
$$\bra 0 | \bar q_2(0) \Gamma [0,x] q_1(x)|V(p)\rangle \quad \mbox{or}\quad
\bra 0 | \bar q_2(0) [0,vx]\Gamma G^a_{\mu\nu}(vx)\lambda^a/2 [vx,x]
q_1(x)|V(p)\rangle.
$$
$x$ is light-like or close to light-like and the light-cone expansion
is an expansion in $x^2$; $v$ is a number between 0 and 1 and $\Gamma$
a combination of Dirac matrices. The expressions $[0,x]$ etc.\ denote
Wilson lines that render the matrix elements, and hence
the DAs, gauge-invariant. One usually works in the convenient
Fock-Schwinger gauge $x^\mu A^a_\mu(x)\lambda^a/2 = 0$, where all Wilson
lines are just $\bf 1 $;
we will suppress them from now on. 

The DAs are formally
ordered by twist, i.e.\ the difference between spin and dimension of
the corresponding operators. In this paper we take into account 2-
and 3-particle DAs of twist-2, 3 and 4.
The classification scheme of vector meson DAs is more involved than
that for pseudoscalars; it has been studied in detail in
Ref.~\cite{wavefunctions}. One important point is the distinction
between chiral-even and chiral-odd operators, i.e.\ those with an odd
or even number of $\gamma_\mu$-matrices. In the limit of massless quarks
the DAs associated with these operators form two completely separate
classes that do not mix under a change of $\mu_{\rm IR}$.
One more important parameter is the
polarisation-state of the meson, longitudinal ($\parallel$) or
transverse ($\perp$), which
helps to classify twist-2 and 3 DAs. Up to twist-4 accuracy, we have
the following decomposition of chiral-even 2-particle 
DAs \cite{wavefunctions}: 
\begin{eqnarray}
\langle 0|\bar q_2(0) \gamma_\mu q_1(x)|V(P,\lambda)\rangle 
 &=& f_V m_V \Bigg\{
\frac{e^{(\lambda)}z}{Pz}\, P_\mu \int_0^1 du \,e^{-i u Pz}
\Big[\phi_\parallel(u)
+\frac{m^2_V x^2}{16}  {\mathbb A}_\parallel(u) + O(x^4)\Big]
\nonumber\\
&&{}+\left(e^{(\lambda)}_\mu-P_\mu\frac{e^{(\lambda)}z}{Pz}\right)
\int_0^1 du\, e^{-i u Pz} \,\left( g_\perp^{(v)}(u) + O(x^2)\right)
\nonumber\\
\lefteqn{-\frac{1}{2}z_\mu \frac{e^{(\lambda)}z}{(pz)^2} m^2_V \int_0^1 du 
\, e^{-i u pz}\,\left( g_3(u) + \phi_\parallel(u) - 
2 g_\perp^{(v)}(u) + O(x^2)\right)
\Bigg\},}\hspace*{2cm}
\label{eq:OPEvector}\\
\langle 0|\bar q_2(0) \gamma_{\mu} \gamma_{5} 
q_1(x)|V(P,\lambda)\rangle 
&=& -\frac{1}{4}\,f_{V}
m_{V} \epsilon_{\mu}^{\phantom{\mu}\nu \alpha \beta}
e^{(\lambda)}_{\nu} p_{\alpha} z_{\beta}
\int_{0}^{1} \!du\, e^{-i u p  z} \left( g_\perp^{(a)}(u) +
O(x^2)\right),\label{eq:OPEaxial}
\end{eqnarray}
and for the chiral-odd ones:
\begin{eqnarray}
\lefteqn{\langle 0|\bar q_2(0) \sigma_{\mu \nu}
q_1(x)|V(P,\lambda)\rangle  = 
 i f_{V}^{T} \left[ (e^{(\lambda)}_{\mu}P_\nu -
e^{(\lambda)}_{\nu}P_\mu )
\int_{0}^{1} \!du\, e^{-i u P z}
\Bigg[\phi_{\perp}(u) +\frac{m_V^2x^2}{16} 
{\mathbb A}_\perp(u)\Bigg] \right.}\hspace*{2cm}\nonumber\\
& &\hspace*{-10pt} {}+ (p_\mu z_\nu - p_\nu z_\mu )
\frac{e^{(\lambda)} z}{(p z)^{2}}
m_{V}^{2}
\int_{0}^{1} \!du\, e^{-i u p z}
\left(h_\parallel^{(t)}(u)-\frac{1}{2}\, \phi_\perp(u) - \frac{1}{2}\,
h_3(u) + O(x^2) \right)
\nonumber
\end{eqnarray}
\begin{equation}
\hspace*{-10pt} \left.{}+ \frac{1}{2}
(e^{(\lambda)}_{ \mu} z_\nu -e^{(\lambda)}_{ \nu} z_\mu)
\frac{m_{V}^{2}}{p  z}
\int_{0}^{1} \!du\, e^{-i u p z} \left(h_3(u)-\phi_\perp(u) +
O(x^2)\right) \right],
\label{eq:OPE2}
\end{equation}
\begin{equation}
\langle 0 | \bar q_2(0) q_1(x) | V(P,\lambda)\rangle =
  \frac{i}{2}\,f_V^T
\left(e^{(\lambda)} z\right) m_V^2 \int_0^1 du\, e^{-i u pz} \left(
h_\parallel^{(s)}(u) + O(x^2)\right).\hspace*{2cm}
\label{eq:OPEx}
\end{equation}
The relevant 3-particle DAs are defined in App.~\ref{app:A}.

Note that we distinguish between light-like vectors $p,z$ with
$p^2=0=z^2$ and the vectors $P,x$ with $P^2=m_V^2$ and $x^2\neq 0$;
explicit relations between these vectors are given in App.~\ref{app:A}. 
The DAs are dimensionless functions of $u$ and
describe
the probability amplitudes to find the vector meson $V$ in a state with minimal
number of constituents ---
 quark and antiquark --- which carry
momentum fractions $u$ (quark) and $1-u$ (antiquark), respectively.
The eight DAs $\phi=\{\phi_{\parallel,\perp},
g_\perp^{(v,a)},h_\parallel^{(s,t)},h_3,g_3\}$ are normalized as
\begin{equation}
\int_0^1\!du\, \phi(u) =1.
\label{eq:norm}
\end{equation}
The nonlocal operators on the left-hand side are
 renormalized at scale $\mu$, so that
 the distribution amplitudes depend on $\mu$ as well. This
dependence can be calculated in perturbative QCD; we will come back to
 that point   below.
                                                                      
The vector and tensor  decay constants $f_V$ and $f_V^T$ featuring in
Eqs.~(\ref{eq:OPEvector}) and (\ref{eq:OPEx}) are  defined
as
\begin{eqnarray}
\langle 0|\bar q_2(0) \gamma_{\mu}
q_1(0)|V(P,\lambda)\rangle & = & f_Vm_V
e^{(\lambda)}_{\mu},
\label{eq:fr}\\
\langle 0|\bar q_2(0) \sigma_{\mu \nu}
q_1(0)|V(P,\lambda)\rangle &=& i f_V^{T}(\mu)
(e_{\mu}^{(\lambda)}P_{\nu} - e_{\nu}^{(\lambda)}P_{\mu});
\label{eq:frp}
\end{eqnarray}
 numerical values are given in Tab.~\ref{tab:fV}. $f_V^T$
depends on the renormalization scale as 
$$
f_V^T(Q^2) = L^{C_F/\beta_0} f_V^T(\mu^2)
$$
with $L = \alpha_s(Q^2)/\alpha_s(\mu^2)$ and $\beta_0 = 11-2/3 n_f$,
$n_f$ being the number of flavors involved.

The DAs as defined above do actually not all correspond to matrix
elements of operators with definite twist: 
$\phi_{\perp,\parallel}$ are of twist-2,
$h_{\parallel}^{(s,t)}$ and $g_\perp^{(v,a)}$ contain a mixture of
twist-2 and 3 contributions and ${\mathbb A}_{\perp,\parallel}$, $h_3$
and $g_3$ a mixture of twist-2, 3 and 4 contributions. Rather than as
matrix elements of operators with definite twist, the DAs
are defined as matrix elements of operators built from fields with a
fixed spin-projection onto the light-cone. For quark fields, the
possible spin projections are $s=\pm 1/2$ and the corresponding
projection operators $P_+=1/(2pz)\, \pslash\zslash$ and $P_- = 1/(2pz)\,
\zslash\pslash$. Fields with fixed spin-projection have a definite
conformal spin, given by $j=1/2 (s+\mbox{canonical mass dimension})$,
and composite operators built from such fields can be expanded in
terms of increasing conformal spin.\footnote{For a more detailed
  discussion we refer to the first reference in
  \cite{wavefunctions} and to Ref.~\cite{VBreview}.} The
expansion of the corresponding DAs, suitably dubbed
conformal expansion, is one of the primary tools in the analysis of
meson DAs, and together with the use of the QCD
equations of motion it allows one to parametrize the plethora of
 2- and 3-particle DAs in terms of a manageable number of
independent hadronic matrix elements.  DAs defined as matrix elements
of operators with definite twist, on the other hand, do not have a 
well-defined conformal expansion \cite{lazar}, and
this is the reason why we prefer the above definitions. In an
admittedly rather sloppy way we will from now on refer to $g_\perp^{(v,a)},
h_\parallel^{(s,t)}$ as twist-3 DAs and to $h_3,g_3,{\mathbb
  A}_{\perp,\parallel}$ as twist-4 DAs. A more detailed
discussion of the relations between the different DAs is given in  
App.~\ref{app:A}; the upshot is that
the 18 twist-2, 3 and 4 DAs we shall take into account 
can be paramatrized, to NLO in the conformal expansion, in
terms of 10 hadronic matrix elements, most of which give only
tiny contributions to the LCSRs for form factors. 

For the leading twist-2 DAs $\phi_{\parallel,\perp}$ in particular,
the conformal expansion goes in terms of Gegenbauer polynomials:
\begin{equation}
\phi(u,\mu^2) = 6 u (1-u) \left( 1 + \sum\limits_{n=1}^\infty
  a_{n}(\mu^2) C_{n}^{3/2}(2u-1)\right).
\end{equation}
The first term on the right-hand side, $6 u (1-u)$, is referred to as
asymptotic DA; as the anomalous dimensions of $a_n$ are positive,
$\phi$ approaches the asymptotic DA in the limit $\mu^2\to\infty$.
The usefulness of this expansion manifests itself in the fact that,
to leading logarithmic accuracy, the (nonperturbative) Gegenbauer
moments $a_n$ renormalize multiplicatively with
\begin{equation}
a_n(Q^2)  =  L^{\gamma_n/(2\beta_0)}\, a_n(\mu^2)
\end{equation}
with $L= \alpha_s(Q^2)/\alpha_s(\mu^2)$.
The anomalous dimensions $\gamma^{\parallel,\perp}_n$ are given by
\begin{eqnarray}
\gamma^\parallel_n &=&  8C_F \left(\psi(n+2) + \gamma_E - \frac{3}{4} -
  \frac{1}{2(n+1)(n+2)} \right),\\
\gamma^\perp_n &=&  8C_F \left(\psi(n+2) + \gamma_E - 1 \right)\label{17}
\end{eqnarray}
with $\psi(n+1) = \sum_{k=1}^n 1/k - \gamma_E$. 
As the contributions from different comformal spin do not mix under
renormalization, at least to leading logarithmic accuracy, one can construct
models for DAs by truncating the expansion
at a fixed order. Despite the absence of any ``small parameter'' in
that expansion, the truncation is justified inasmuch as one is
interested in physical amplitudes rather than the DA itself. If we write
$${\rm amplitude~}= \int_0^1 du\,\phi(u) T(u),
$$
then, assuming that $T$ is a regular function of $u$, i.e.\ with no
(endpoint) singularities, the highly
oscillating behavior of the Gegenbauer polynomials suppresses
contributions from higher orders in the conformal expansion. 
Even for a function $T$ with a mild endpoint singularity, 
for instance $T=\ln u$, we find, using the
generating function of the Gegenbauer polynomials,
$$\int_0^1 du\,\phi(u) T(u) = -\frac{5}{6}\,a_0 + \sum_{n=1}^\infty
  \frac{(-1)^{n-1} }{n(n+3)}\,3a_n.$$
This  result indicates that, assuming the $a_n$ fall off in $n$,
  which, as we shall see seen in Sec.~\ref{sec:4.3}, is indeed the case,  
even a truncation after the first few terms should
  give a reasonable approximation to the full amplitude. A more
  thorough discussion of the convergence of the conformal expansion
  for physical amplitudes can be found in Ref.~\cite{angi}. 
The major shortcoming of models based on the truncation of the
  conformal expansion is the fact that the
  information available on the actual values of the $a_n$ (and in particular
  their analogues in 3-particle DAs) is, to put it mildly,
 scarce. We therefore use
   truncated models only for DAs whose contribution to the LCSRs is
  small as is the case for all 3-particle DAs and the twist-4 DAs;
  explicit formulas are given in App.~\ref{app:A}.
  All contributions due to or induced by twist-2 DAs, on the other
  hand, are treated as described in Sec.~\ref{sec:4.3}. 

The major difference between the analysis of LCSRs for $B\to\,$vector
meson form factors and that of $B\to\,$pseudoscalar form factors
presented in \cite{BZ} is probably the identification of a suitable
parameter by which to order the relative weight of different contributions
to the sum rules. For $B\to\,$pseudoscalar form factors, the standard
classification in terms of increasing twist proved to be
suitable, as the chiral parities of the twist-2 DA and 2-particle twist-3
DAs are different, so that contribution of the latter to the LCSRs is
suppressed by a factor $m_\pi^2/(m_u+m_d)/m_b$. In addition, the 
admixture of twist-2 matrix elements to twist-3 DAs
and of twist-2 and 3 matrix elements to twist-4 DAs is small and
moreover vanishes in the chiral limit $m_\pi\to 0$.
For vector mesons, the
situation is more complex: for instance, both the twist-2 DA
$\phi_\perp$ and the twist-3 DA $g_\perp^{(v)}$ contribute at the same
order to the form factors $A_2$ and $A_0$, in the combination 
$\phi_\parallel - g_\perp^{(v)}$. Naive twist-counting is evidently 
not very appropriate for classifying the relative size of
contributions of different DAs to the form factors. Instead, we 
decide to classify the relevance of
contributions to the LCSRs not by twist, but by a parameter
$\delta\propto m_V$. The precise definition of $\delta$ depends on the
kinematics of the process; to leading order in an expansion in
$1/m_b$, however, one finds $\delta_{\rm HQL} = m_V/m_b$. 
The numerical analysis of the
LCSRs does indeed display a clear suppression of terms in 
$O(\delta)$ and higher, which suggests
the following  classification of 2-particle DAs:
\begin{itemize}
\item $O(\delta^0):$ $\phi_\perp$;
\item $O(\delta^1):$ $\phi_\parallel,g_\perp^{(v,a)}$;
\item $O(\delta^2):$ $h_\parallel^{(s,t)},h_3,{\mathbb A}_\perp$;
\item $O(\delta^3):$ $g_3,{\mathbb A}_\parallel$.
\end{itemize}
We treat $\delta$ as expansion parameter of the light-cone
expansion and shall combine it with the perturbative QCD expansion in
$\alpha_s$ to obtain a second order expression for the correlation
functions (\ref{eq:corr}); terms in $\delta^3$ are dropped.

\section{Calculation of the Correlation Functions}\label{sec:3}

As we have seen in Sec.~\ref{sec:2}, LCSRs for form factors are
  extracted from the correlation function of the corresponding weak
  current with the pseudoscalar current $j_b = m_b \bar q' i\gamma_5
  b$, evaluated between the vacuum and the vector meson. In this
  section we describe the calculation of these correlation functions
 to second order in $\alpha_s$ and $\delta$.

The relevant correlation functions are defined as
\begin{eqnarray}
\lefteqn{i\int d^4x e^{iqx} \langle V(p)|T(V-A)_\mu(x)
j_b^\dagger(0)|0\rangle =}\hspace*{0.5cm}\nonumber\\ 
& = & -i \Gamma_0 e^*_\mu + i \Gamma_+(e^* q)\, (q+2p)_\mu +
i \Gamma_- \,(e^* q)\, q_\mu + \Gamma_V
\epsilon_\mu^{\phantom{\mu}\alpha\beta\gamma} e^*_\alpha q_\beta
p_\gamma\,,\label{15}
\end{eqnarray}
\begin{eqnarray}
\lefteqn{i\int d^4x e^{iqx} \langle V(p)|T [\bar q\sigma_{\mu\nu}
  \gamma_5 b](x)
 j_b^\dagger(0)|0\rangle =}\hspace*{0.5cm}\nonumber\\
& = &  {\cal A} \{e^*_\mu (2p+q)_\nu - e^*_\nu
(2p+q)_\mu\}
 - {\cal B}\{e^*_\mu q_\nu - e^*_\nu q_\mu\} - 2
{\cal C} (e^* q) \{p_\mu q_\nu - q_\mu p_\nu\}.\label{x}
\end{eqnarray}
The definitions of $\Gamma^{\pm}$ and ${\cal C}$ differ from those used
in Ref.~\cite{BB98} by a factor $pq$; we shall come back to this
point below. In this section we describe the calculation of the
contributions of 2-particle DAs to the above correlation functions;
those of 3-particle DAs are calculated in App.~\ref{app:B}.
 
\begin{figure}[tb]
$$\epsfxsize=0.55\textwidth\epsffile{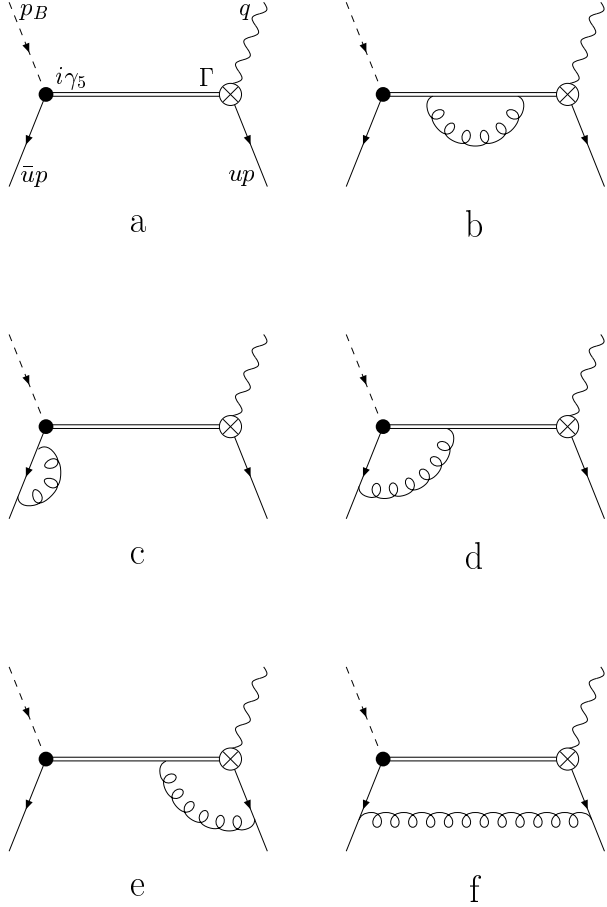}$$
\caption[]{Diagrams for 2-particle correlation functions. $\Gamma$
  is the weak interaction vertex. The light-quark self-energy diagrams
  of type c give purely divergent contributions in $1/\epsilon_{\rm IR} -
  1/\epsilon_{\rm UV}$.}\label{fig:1}
\end{figure}

In light-cone expansion and including only contributions from
2-particle Fock-states of the mesons, each of the seven invariants
$\Gamma_{0,\pm,V}$, ${\cal A,B,C}$ can be written as a
convolution integral of type
\begin{equation}\label{eq:fac}
\Gamma^V = \int \frac{dk_+}{2\pi}\, \phi^V_{ab}(k_+)\, T_{ba}(k_+,p_B^2,q^2)
\end{equation}
with $a,b$ being spinor indices. $p^2=m_V^2$ is set to 0 and 
$k_+$ is the longitudinal momentum of
the quark in the 
vector meson $V$, which is related to the momentum fraction $u$
introduced in Sec.~\ref{sec:2} by $k_+ = u p_+$.\footnote{The
  plus-component of a 4-vector $k^\mu$ is defined as $k_+ =
  (k^0+k^3)/\sqrt{2}$, the minus-component as $k_- =
  (k^0-k^3)/\sqrt{2}$.}
The above factorisation formula implies a complete decoupling of
long-distance QCD effects, encoded in the DA $\phi^V$, and short-distance
effects calculable in perturbation theory, described by
$T$. Factorisation also makes it possible to calculate $T$ in a
convenient way: if it holds, $T$ must be independent of the specific
properties of the external hadron state, 
and one can calculate $\Gamma^V$
 with a particularly simple state that allows a
straight\-for\-ward extraction of the short-distance amplitudes 
$T$.\footnote{This is completely
  analogous to the calculation of Wilson-coefficients in a local 
operator product expansion, which must be independent of the external
states and hence are calculated using any convenient state.} A
convenient choice of the external state is a free quark-antiquark pair
with longitudinal momenta $up$ and $\bar u p$ and spins $s$ and $r$, 
respectively, and DA
\begin{eqnarray*}
\phi^{q_1\bar q_2}_{ab}(k_+) &=& \left.\int dz_- \,e^{-ik_+z_-} 
\langle q_1(u p,s) \bar q_2(\bar u p,r) | (\bar q_1)_{a}(z) [z,0] (q_2)_b(0)
| 0 \rangle\right|_{z_+=0,z_\perp=0}\\
& = & 2\pi\,\bar u^{q_1}_{a}(up,s) v^{q_2}_{b}(\bar up,r)\delta(k_+-up_+),
\end{eqnarray*}
where $\bar u$ and $v$ are the standard fermion spinors.
The $T$ amplitudes, to one loop accuracy, 
are then given directly by the diagrams  shown in
Fig.~\ref{fig:1} with external on-shell quarks with momenta $up$ and
$(1-u)p$, respectively.
The projection onto a specific Dirac structure is done using the general
decomposition
\begin{eqnarray}
(\bar q_1)_a (q_2)_b & = & \frac{1}{4}\, ({\bf 1})_{ba} (\bar q_1 q_2) -
\frac{1}{4} \, (i\gamma_5)_{ba} (\bar q_1 i\gamma_5 q_2) + \frac{1}{4}\,
(\gamma_\mu)_{ba} (\bar q_1 \gamma^\mu q_2) - \frac{1}{4}\,
(\gamma_\mu\gamma_5)_{ba} (\bar q_1 \gamma^\mu \gamma_5 q_2)\nonumber\\
& & {} + \frac{1}{8}\,(\sigma_{\mu\nu})_{ba} (\bar q_1
  \sigma^{\mu\nu} q_2).\label{20A}
\end{eqnarray}
In order to obtain the convolution integrals for
vector mesons, one has to replace the structures $\bar q_1 \Gamma q_2$
in (\ref{20A}) by the appropriate DAs and include 
factors of $e^*z$, $pz$ and
$x^2$ as given in Eqs.~(\ref{eq:OPEvector}) to (\ref{eq:OPEx}). 
The translation of explicit terms in $z_\mu$ 
into momentum space is given by
\begin{equation}\label{deriv1}
z_\mu \to -i \frac{\partial}{\partial(up)_\mu},
\end{equation}
as the outgoing $q_1$ comes with a factor $\exp(iupz)$.
Terms in $1/(p z)$ can be treated by partial integration:
\begin{equation}\label{deriv2}
\frac{1}{p z}\, \phi(u) \to -i \int_0^u dv\,\phi(v) \equiv -i \Phi(u).
\end{equation}
There are no surface terms, as for all the relevant structures $\phi$,
e.g.\ $\phi_\parallel-g_\perp^{(v)}$, one has $\Phi(0)=0=\Phi(1)$. 
A second, approximate way to deal with factors $(e^* z)/(pz)$ is based
on the observation that $(e^* z)$ projects onto the longitudinal
polarisation state of the vector meson, cf.\ Eq.~(\ref{polv}), and that 
in the ultrarelativistic limit $E_V\to\infty$ the longitudinal
polarisation vector is approximately collinear with the meson's momentum:
$$
\epsilon^{(0)}_\mu = \frac{1}{m_V}\left( p_\mu +
O(m_V^2)\right)\quad\Longrightarrow \quad \frac{e^* z}{pz}\to
\frac{1}{m_V}\mbox{~~and~~} \frac{1}{m_V}\to \frac{e^* q}{pq}.
$$
Up to corrections in $m_V^2$, this procedure yields results identical  
with those from partial
integration --- provided that the corresponding DA $\phi$ is normalized to
0. That is:
$$
-\int_0^1 du
\Phi(u)\, e^*_\kappa\,\frac{\partial}{\partial(up)_\kappa} \sim 
\frac{e^*z}{pz} \int_0^1 du \phi(u) \sim \frac{e^* q}{pq} \left(
\int_0^1 du \phi(u) + O(m_V^2)\right), 
$$
where the first relation is valid only if $\phi$ is normalized to 0,
i.e.\ $\Phi(1) = 0$. This is
indeed the case for the mixed-twist structure
$\phi_\parallel-g_\perp^{(v)}$, but does not apply if only the
pure twist-2 DA $\phi_\parallel$ is included, as  
done in \cite{BB98}. In this
case, unphysical singularities in $p_B^2=q^2$ appear in $\Gamma_\pm$
and ${\cal C}$ and have to be factored out. 
This explains the appearance of additional factors $1/(pq)
= 2/(p_B^2-q^2)$ in the correlation functions used in \cite{BB98}. 
In the calculation presented in this paper
we use the prescriptions (\ref{deriv1}) and (\ref{deriv2}) througout
and hence avoid unphysical singularities in $p_B^2$. We have
checked that indeed the singularity structure of all 
seven invariants $\Gamma_{0,\pm,V}$, ${\cal A,B,C}$ is given by a cut
on the real axis for $p_B^2\geq m_b^2$. 

The complete correlation function, including 2-particle DAs 
to $O(\delta^2)$ for the vector current $V$, axialvector current $A$, tensor
current $T$ and scalar current $S$,\footnote{The matrix elements of
  vector mesons over the pseudoscalar current vanish.} can now be written as
\begin{eqnarray}
\Gamma(q^2,p_B^2) &=& 
\sum_{C=V,A,T,S}\int_0^1 du\,{\mathfrak P}_{ab}^C(u;\mu^2)
T_{ba}(u,p_B^2,q^2;\mu^2)\\[-5pt]
{\rm with}\quad {\mathfrak P}^V_{ab} & = & \frac{1}{4}\, f_V m_V
(\gamma^{\alpha})_{ab} \left( -P_\alpha e^*_\beta
 \,\Phi(u)\,\frac{\partial}{\partial(up)_\beta} + e^*_\alpha
g_\perp^{(v)}(u) \right),\\
{\mathfrak P}^A_{ab} & = &
-\frac{i}{16}\,f_Vm_V\,(\gamma^\alpha\gamma_5)_{ab}\,
\epsilon_{\alpha\kappa\lambda
  \beta}\,e^{*\kappa} P^\lambda
g_\perp^{(a)}(u)\,\frac{\partial}{\partial(up)_{\beta}},\\
{\mathfrak P}^T_{ab} & = & -\frac{i}{4}\, f_V^T
(\sigma^{\alpha\beta})_{ab} \left\{ e^*_\alpha P_\beta \left(
\phi_\perp(u) - \frac{1}{16}\,m_V^2 {\mathbb A}_\perp
\,\frac{\partial^2}{\partial(up)_\kappa\partial(up)^\kappa}\right)\right.\\
&&{}\left.+
  m_V^2 P_\alpha \epsilon^*_\gamma
  I_L(u)\,\frac{\partial^2}{\partial(up)_\beta \partial(up)_\gamma} -
  \frac{1}{2}\,m_V^2\,e^*_\alpha H_3(u)
  \,\frac{\partial}{\partial(up)_\beta}\right\},\\
 {\mathfrak P}^S_{ab} & = &
-\frac{1}{8}\,m_V^2 f_V^T (e^*_\alpha)({\bf 1})_{ab}
h_\parallel^{(s)}(u) \,\frac{\partial}{\partial(up)_\beta},\\[-5pt]
{\rm where}\quad 
\Phi(u) & = & \int_0^v dv
\left(\phi_\parallel(v)-g_\perp^{(v)}(v)\right),\nonumber\\
I_L(u) & = & \int_0^u dv \int_0^v dw \left(h_\parallel^{(t)}(w)
-\frac{1}{2}\, \ \phi_\perp(w) - \frac{1}{2}\, h_3(w)\right),\nonumber\\
H_3(u) & = & \int_0^u dv\,(h_3(v)-\phi_\perp(v)).\nonumber
\end{eqnarray}
All these three functions $F(u)$ fulfill $F(0)=1=F(1)$.
                                   
Just to give an example, the tree-contribution is given by
$$
T_{ba}^{\rm tree} = -i\left[\Gamma (\slash{q}+u\slash{P}+m_b)
\gamma_5\right]_{ba}/((q+uP)^2-m_b^2)
$$
with the weak vertex $\Gamma$.

In order for factorisation to hold, two conditions have to be met:
\begin{itemize}
\item[(a)] the long-distance infrared
sensitive parts (IR-sin\-gu\-la\-ri\-ties) in $T$ have to cancel against those 
in the DAs;
\item[(b)] the convolution integral $\int du \phi^{\rm ren}(u) T^{\rm
  ren}(u)$ has to converge. Otherwise factorisation is violated by
  soft end-point singularities.
\end{itemize}
In order to check condition (a), we  
decompose the bare amplitude into finite and
divergent terms as
\begin{eqnarray*}
T^{\rm bare}(u) & = & T^{(0)}(u) + \alpha_s \left( 
T^{(1),{\rm ren}}(u) +
  \frac{1}{\epsilon} \,T^{(1),\rm div}(u)\right).
\end{eqnarray*}
Ultra-violet divergences, which only occur for the
penguin-current, are easily subtracted using the known renormalisation
of the corresponding current:
$$T^{\rm bare}(u) \to T^{\rm bare}(u) - \delta Z_{\rm peng}
T^{(0)}(u).$$
The remaining divergent terms have to cancel against the divergent
parts of the bare DA,
$$\phi^{\rm bare}(u) = \phi^{\rm ren}(u) + \alpha_s
\,\frac{1}{\epsilon}\, \phi^{\rm div}(u)
$$ 
so that
$$\int_0^1 du\left( \phi^{\rm ren}(u) T^{(1),{\rm div}}(u) + \phi^{\rm
  div}(u) T^{(0)}(u) \right) = 0.
$$
$\phi^{\rm div}(u)$ is known explicitly for the twist-2 $\pi$ DA
\cite{pQCD} and
coincides with that for $\phi_\parallel$, but to the best of our
knowledge has not yet been calculated for $\phi_\perp$. Alternatively,
one can check the cancellation of divergences order by order in the
conformal expansion of the DAs, cf.\ Sec.~\ref{sec:2} and
App.~\ref{app:B}, with\footnote{We use dimensional regularisation with
$D=4+2\epsilon$.} 
$$a_n^{\parallel,\perp,{\rm bare}} = a_n^{\parallel,\perp} \left( 1 +
\frac{\alpha_s}{4\pi}\,
\frac{\gamma_n^{\parallel,\perp}}{2}\,\frac{1}{\epsilon} \right).
$$
We find that all $1/\epsilon$ terms cancel as required. 

As for condition (b), we also find that all $T$ are
regular at the endpoints, so that there are no endpoint singularities
in the convolution.

As an
interesting by-product, we also find the following fixed-order 
evolution-equations of the first
inverse moment of the DAs:
\begin{eqnarray}
\int_0^1 du\,\frac{\phi_\parallel(u,\mu_2^2)}{u} &=& 
\int_0^1 du\,\frac{\phi_\parallel(u,\mu_1^2)}{u}\left\{ 1 +
a_s\,\ln\,\frac{\mu_2^2}{\mu_1^2} \left( 3 + 2 \ln u\right)\right\},\nonumber\\
\int_0^1 du\,\frac{\phi_\perp(u,\mu_2^2)}{u} &=& 
\int_0^1 du\,\frac{\phi_\perp(u,\mu_1^2)}{u}\left\{ 1 +
2 a_s\,\ln\,\frac{\mu_2^2}{\mu_1^2} \left( 2 + \frac{\ln
  u}{1-u}\right)\right\}.\label{25}
\end{eqnarray}
These equations allow one to calculate the change of that inverse
moment directly for a given DA without having to calculate the
Gegenbauer-moments in an intermediate step. 
The first of these relations can also be obtained
from the known one-loop evolution kernel of the $\pi$ twist-2 DA
\cite{pQCD}, whose anomalous dimensions coincide with those of
$\phi_\parallel$; the relation for $\phi_\perp$ is new.

As we shall see in the next section, the LCSRs do actually not
involve the full correlation functions, but only their imaginary
parts in $p_B^2$. As in Ref.~\cite{BZ} we take the imaginary part only after
calculating the convolution integral, which results in closed and
comparatively simple, albeit lengthy expressions. 
The distribution amplitudes $\phi_{\perp,\parallel}$,
$g_\perp^{(v,a)}$ are given by
their respective conformal expansions, which we truncate at $a_9$. 
As discussed in Sec.~\ref{sec:2}, the effective expansion parameter of
the light-cone expansion is $\delta$, so that the correlation
function is expanded in both $\delta$ and $\alpha_s$. 
We combine both expansions and include terms up to second order, i.e.\
$O(\delta^{0,1,2}\alpha_s^0)$ and $O(\delta^{0,1}\alpha_s^1)$, but
  drop $O(\delta^2\alpha_s^1)$.\footnote{Terms of
    $O(\delta^0\alpha_s^2)$ are not included, either.}
 A list of the included terms is given in
Tab.~\ref{tab:1}. Note that we have not calculated the radiative
corrections to the contributions from the 3-particle twist-3 DAs
$\cal V,A$ as they are expected to be very small. This follows in part
from the observation that $O(\alpha_s)$ terms in the corresponding 
twist-3 matrix elements do also show up in the $O(\alpha_s)$
corrections to $g_\perp^{(v,a)}$ and are very small numerically. 
\begin{table}
\renewcommand{\arraystretch}{1.2}\addtolength{\arraycolsep}{3pt}
$$\begin{array}{|llcc|}\hline
& {\rm DA} & O(\alpha_s^0) & O(\alpha_s)\\\hline
\mbox{\rm twist-2:} & \phi_\perp & \delta^0,\delta^2 & \delta^0\\
& \phi_\parallel & \delta\phantom{^{0}} & \delta\phantom{^{0}}\\
\mbox{\rm twist-3:} & g^{(a)}_\perp,g_\perp^{(v)} 
& \delta\phantom{^{0}}&  \delta^{(*)}\\
& h_\parallel^{(s)},h_\parallel^{(t)} & \delta^2 & -\\
& {\cal V,A}\mbox{~(3-part.\ DAs)} & \delta\phantom{^{0}} & -\\
& {\cal T}\mbox{~(3-part.\ DA)} & \delta^2 & -\\
\mbox{\rm twist-4:} & h_3, {\mathbb A}_\perp & \delta^2 & - \\
& \mbox{\rm chiral-odd 3-part.\ DAs} & \delta^2 & - \\\hline
\end{array}$$
\renewcommand{\arraystretch}{1}\addtolength{\arraycolsep}{-3pt}
\caption[]{Contributions included in the calculation of the
  correlation functions (\ref{15}) and (\ref{x}). $\delta\propto
  m_V$ is the
  effective expansion parameter of the light-cone expansion; we include
  contributions up to second order in $\delta$ and $\alpha_s$; those
  marked by $(*)$ are new.}\label{tab:1}
\end{table}

Depending on the specific weak vertex and projection onto the DAs, some
diagrams contain traces with an odd number
of $\gamma_5$, which leads to ambiguities
when naive dimensional regularisation with
anticommuting $\gamma_5$ is used. We solve this problem by using Larin's
prescription for dealing with $\gamma_5$ \cite{Larin} and replace, whenever
necessary,  ($a_s
= C_F \alpha_s/(4\pi)$)
\begin{eqnarray*}
\gamma_\mu\gamma_5 & \to & \left(1- 4 a_s\right) \frac{i}{3!}\,\epsilon_{\mu
  \nu_1\nu_2\nu_3} \,\gamma^{\nu_1}\gamma^{\nu_2}\gamma^{\nu_3},\\
\gamma_5 & \to & \left(1- 8 a_s\right) \frac{i}{4!}\,\epsilon_{\nu_1
  \nu_2\nu_3\nu_4} \,\gamma^{\nu_1}\gamma^{\nu_2}\gamma^{\nu_3}
\gamma^{\nu_4},
\end{eqnarray*}
\begin{eqnarray*}
\sigma_{\mu\nu}\gamma_5 &\to& -\left(1- 0 a_s\right) \frac{i}{2}
  \,\epsilon_{\mu\nu\alpha\beta} \sigma^{\alpha\beta}.
\end{eqnarray*}
Note that we use the Bjorken/Drell convention for the $\epsilon$
tensor with $\epsilon_{0123}=+1$.
For the special case of the axial-vector form factors and the
projection onto the DA $g_\perp^{(a)}$, one can implement Larin's prescription
by rewriting either the weak vertex or the $B$ vertex. We have checked
that we obtain the same result in both cases. One might also think of
``Larinizing'' the projection operator onto the DA; the corresponding
finite renormalisation will be $u$-dependent due to the nonlocality
of the current and is yet unknown.

\section{Numerics}\label{sec:4}

This section is the heartpiece of our paper, in which 
we derive the sum rules for $B\to V$ form factors and
obtain numerical results. The section is organised as follows: in
Sec.~\ref{sec:4.1} we derive the LCSR for one of the seven form
factors, $V$. In Sec.~\ref{sec:4.2} we give values for most of the
needed hadronic input parameters and explain how to determine the sum rule
specific parameters, i.e.\ the Borel parameter $M^2$ and the continuum
threshold $s_0$. We also calculate $f_{B_d}$ and $f_{B_s}$, which are necessary
ingredients in the LCSRs. In Sec.~\ref{sec:4.3} we motivate the need
for and introduce models of the twist-2 DAs $\phi_{\perp,\parallel}$.
In Sec.~\ref{sec:4.4} we calculate the form factors at $q^2=0$
and discuss their uncertainties. 
In Sec.~\ref{sec:4.5} we present the form factors for
central input values of the parameters and provide a simple
parametrization valid in the full kinematical regime of $q^2$.
The results for $q^2=0$ are collected in Tab.~\ref{tab:basic}, central 
results for arbitrary $q^2$ in
Tab.~\ref{tab:fits}. 

\subsection{The Sum Rules}\label{sec:4.1}

With explicit expressions for the correlation functions in hand, we
are now in a position to derive the LCSRs for the form factors. Let us
choose $V(q^2)$ for a $B_q$ transition as example. 
The corresponding correlation function is $\Gamma_V$ as
defined in Eq.~(\ref{15}). The basic idea is to express $\Gamma_V$ in
two different ways, as dispersion relation of the expression obtained
in light-cone expansion on one hand, and as dispersion relation in
hadronic contributions on the other hand. Equating  both
representations one obtains a light-cone sum rule for $V$.
One side of the equation is hence the light-cone expansion result
\begin{equation}\label{xy}
\Gamma_V^{\rm LC}(p_B^2,q^2) = \int_{m_b^2}^\infty ds\,\frac{\rho^{\rm
    LC}_V(s,q^2)}{s-p_B^2}\,,
\end{equation}
with $\pi \rho^{\rm LC}_V(s,q^2)={\rm Im}[\Gamma_V^{\rm LC}]$, which 
has to be compared to the physical correlation function
that  also features a cut in $p_B^2$, starting at $m_B^2$:
\begin{equation}
\Gamma_V^{\rm phys}(p_B^2,q^2) = \int_{m_B^2}^\infty ds\,\frac{\rho^{\rm
    phys}_V(s,q^2)}{s-p_B^2}\,;
\end{equation}
the spectral density is given by hadronic contributions and reads
\begin{equation}\label{xz}
\rho^{\rm phys}_V(s,q^2) = f_{B_q} m_B^2 \,\frac{2V(q^2)}{m_B+m_V}\,
\delta(s-m_B^2) +
\rho^{\rm\scriptstyle higher-mass~states}_+(s,q^2).
\end{equation}
Here $f_{B_q}$ is the $B_q$ meson decay constant defined as
\begin{equation}\label{fb}
\langle 0 |\bar q \gamma_\mu\gamma_5 b | B\rangle = if_{B_q} p_\mu
\quad{\rm or} \quad (m_b+m_q) \langle 0 |\bar q i\gamma_5 b | B\rangle
= m_B^2 f_{B_q}.
\end{equation}
To obtain a light-cone sum rule for $V$, one equates the two
expressions for $\Gamma_V$ and uses quark-hadron duality to approximate
\begin{equation}\label{dunno}
\rho^{\rm\scriptstyle higher-mass~states}_V(s,q^2) \approx \rho_V^{\rm
  LC}(s,q^2)\Theta(s-s_0),
\end{equation}
where $s_0$, the so-called continuum threshold is a parameter to be
  determined within the sum rule approach itself. In principle one
    could now write a sum rule
$$\Gamma_V^{\rm\scriptstyle phys}(p_B^2,q^2) = \Gamma_V^{\rm LC}(p_B^2,q^2)$$
and extract  $V$. However, in order to suppress the impact
of the approximation (\ref{dunno}), one subjects
 both sides of the equation to a Borel
  transformation
$$\frac{1}{s-p_B^2}\to
\hat{B}\,\frac{1}{s-p_B^2} = \frac{1}{M^2}\, \exp(-s/M^2)$$
which ensures that contributions from higher-mass states be sufficiently
suppressed and improves the convergence of the OPE.
We then obtain
\begin{equation}\label{srx}
e^{-m_B^2/M^2}  m_B^2f_{B_q}\;\frac{2V(q^2)}{m_B+m_V} = 
\int_{m_b^2}^{s_0} ds\, e^{-s/M^2}
\rho_V^{\rm LC}(s,q^2). \quad
\end{equation}
This is the final sum rule for $V$ and explains why, as announced in
the previous section, only the imaginary part of the correlation
function is needed. Expressions for the other
form factors are obtained analogously. The task is now to find sets of
parameters $M^2$ (the Borel parameter) and $s_0$ (the continuum
threshold) such that the resulting form factor does not
depend too much on the precise values of these parameters; in
addition the continuum contribution, that is the part of the
dispersive integral from $s_0$ to $\infty$, which has been subtracted
from both sides of (\ref{srx}), should not be too large, say less than
30\% of the total dispersive integral.

\begin{table}
\addtolength{\arraycolsep}{3pt}\renewcommand{\arraystretch}{1.3}
$$
\begin{array}{|l|cccc|}
\hline
V & \rho & \omega & K^* & \phi \\\hline
f_V [{\rm MeV}] & 205\pm 9\phantom{0} & 195\pm 3\phantom{0} & 
217 \pm 5\phantom{0} & 231 \pm 4\phantom{0}\\
f_V^T(1\,{\rm GeV}) [{\rm MeV}] & 160\pm 10 & 145 \pm 10 & 170\pm 10 & 200 \pm
10\\
f_V^T(2.2\,{\rm GeV}) [{\rm MeV}] & 147\pm10 & 133\pm10 & 156\pm10 & 183\pm10 \\\hline
\end{array}
$$
\addtolength{\arraycolsep}{-3pt}\renewcommand{\arraystretch}{1}
\caption[]{Values of the vector meson couplings. $f_V$ is
  extracted from experiment, $f_V^T$ from QCD sum rules for 
$f_V^T/f_V$, cf.\ Ref.~\cite{SU(3)breaking}.}\label{tab:fV}
\end{table}

\subsection{Hadronic Input Parameters}\label{sec:4.2}

After having derived the LCSRs for the form factors, the next step is
to fix the parameters on which they depend. These are the
decay constants of the $B_q$ and $B_s$ meson, $f_{B_q}$ and
$f_{B_s}$, the couplings $f^{(T)}_V$ of the vector mesons, introduced
in Sec.~\ref{sec:2}, the meson DAs, the quark masses $m_b$ and $m_s$,
$\alpha_s$ and the factorisation scale $\mu_{\rm IR}$,
and, finally, the sum-rule specific parameters $M^2$ and $s_0$. 

The $f_V$ are known from experiment and are collected in
Tab.~\ref{tab:fV}. The $f_V^T$, on the other hand, are not that easily
accessible in experiment and hence have to be determined from
theory. For internal consistency, we determine these parameters from
QCD sum rules for the ratio $f_V^T/f_V$, as explained in
Ref.~\cite{SU(3)breaking}. The results are collected in
Tab.~\ref{tab:fV}, too. $f_\rho^T$ had already been determined
earlier in Ref.~\cite{BB96}; the result agrees with that in
Tab.~\ref{tab:fV}. The ratios $f_V^T/f_V$ have also been determined
from lattice \cite{lattfT} and  agree with ours within errors.
Meson DAs are discussed in the next subsection.

The $b$ quark mass entering our formulas is the
one-loop pole mass $m_b$ for which we use  $m_b = (4.80\pm
0.05)\, \text{GeV}$, cf.\ Tab.~6 in Ref.~\cite{LCSRs:reviews}. 
$m_s$, on the other
hand, is the $\overline{\text{MS}}$ running  mass, 
$\overline{m}_s(2\,\text{GeV}) =
100\,\text{MeV}$, which is an average of two recent lattice
determinations \cite{mslatt}; the uncertainty in $m_s$ has only a
minor impact on our results. As for the strong coupling, we take 
$\alpha_s(m_Z)=0.118$ and use NLO evolution to evaluate it at lower
scales. All scale-dependent quantities 
are evaluated at the factorisation scale $\mu_{\rm IR}$ which
separates long- from short-distance physics. The only exception are
the form factors $T_i$, which also depend on an ultraviolet scale
$\mu_{\rm UV}$
which is set to $m_b$. We choose $\mu_{\rm IR}
= \sqrt{m_B^2-m_b^2} = 2.2\,\text{GeV}$ as reference scale; 
a variation of $\mu_{\rm IR}$ by $\pm 1\,{\rm GeV}$ has only small
impact on the final results.

The remaining parameters are $f_{B_{q,s}}$, $M^2$ and $s_0$. $f_{B_{q,s}}$ has
been determined from both lattice and QCD sum rule calculations. The
state of the art of the former are unquenched NRQCD simulations with $2+1$
light flavors, yielding $f_{B_s} = (260\pm 30)\,\text{MeV}$
\cite{fBlatt}, which is
slightly larger than the 2003 recommendation $f_{B_s} = (240\pm
35)\,\text{MeV}$ \cite{kronfeld}. For $f_{B_d}$, it is difficult to
find any recent numbers, the consensus being that more
calculations at smaller quark masses are needed in order to bring
the extrapolation to physical $m_{u,d}$ under sufficient control \cite{fBlatt}.
As for QCD sum rules, both $f_{B_d}$ and $f_{B_s}$ have been determined to
$O(\alpha_s^2)$ accuracy: $f_{B_d} = (208\pm 20)\,\text{MeV}$ and
$f_{B_s} = (224\pm 21)\,\text{MeV}$ \cite{fBSR}, in agreement with
lattice determinations. The impact of $O(\alpha_s^2)$ corrections on
$f_{B_{q,s}}$ is nonnegligible. As the diagrams responsible for these
corrections, for instance $B$ vertex corrections, are precisely the
same that will enter LCSRs at $O(\alpha_s^2)$, we proceed from the
assumption that these corrections 
will tend to cancel in the ratio
(correlation~function)/$f_B$. We hence evaluate $f_{B_{d,s}}$ from a QCD
sum rule to $O(\alpha_s)$ accuracy, which 
reads \cite{SRfB}:\footnote{The contribution of
  the gluon condensate
is not sizable and we therefore neglect it.}
\begin{eqnarray}\label{eq:aa}
f_{B_q}^2 m_B^2 e^{-m_B^2/M^2} &=& \int_{m_b^2}^{s_0} ds\,\rho^{\rm pert}(s)
e^{-s/M^2} + C_{\bar qq} \bra\bar q q \ket + C_{\bar q Gq} 
\bra\bar q \sigma gG q \ket
\equiv \int_{m_b^2}^{s_0} ds\,\rho^{\rm tot}(s)
e^{-s/M^2}.\nonumber\\[-0.5cm]
\end{eqnarray}
Here $ \bra\bar q q \ket$ and $\bra\bar q \sigma gG q \ket$ are the
the quark and mixed condensate, respectively, for which we use the 
following numerical values at $\mu=1\,{\rm GeV}$: 
\begin{equation}\label{eq:conds}\bra\bar q
q\ket = -(0.24\pm0.01)^3\,\text{GeV}^3 \quad\mbox{and} \quad 
{\bra \bar q \sigma gG q \ket}= 0.8\,\text{GeV}^2\bra
\bar q q \ket.
\end{equation}
The $C$ are perturbative Wilson coefficients multiplying the
condensates. $C_{\bar qq}$ is known to $O(\alpha_s)$ accuracy
\cite{thesis,fBSR}, $C_{\bar q Gq}$ at tree-level.

The criteria for choosing $M^2$ and $s_0$ in the above sum
rule are very similar to those to be used for the LCSRs. Ideally, if the
correlation function were known exactly, the sum rule would be
independent of $M^2$. In practice it isn't, but ``good'' sum rules,
plotted as function of $M^2$,
still exhibit a flat extremum. We hence require the existence of such
an extremum in $M^2$ and evaluate the sum rule precisely at that point. This
eliminates $M^2$ as independent parameter and leaves us with $s_0$. As
already mentioned after Eq.~(\ref{srx}), the purpose of the Borel
transformation is to enhance the contribution of the ground state to
the physical spectral function with
respect to that of higher states. We hence
require that that continuum contribution, that is the integral over
$\rho^{\rm tot}(s)$ for $s>s_0$, must not be too large. To be
specific, we require 
$$\left(\int_{s_0}^\infty ds \rho^{\rm tot}(s) e^{-s/M^2}\right)/\left(
\int_{m_b^2}^\infty ds \rho^{\rm tot}(s) e^{-s/M^2}\right) < 30\%.$$
This puts a lower bound on $s_0$. The larger $s_0$, the smaller $M^2$,
the position of the minimum, and the larger nonperturbative
contributions to (\ref{eq:aa}). As the condensates are meant to yield
small nonperturbative corrections, but blow up at small $M^2$,
requiring the nonperturbative corrections to be not too large puts an
upper bound on $s_0$. For $f_{B_{q,s}}$, we require the highest term
in the condensate expansion, the mixed condensate, to contribute less
than $10\%$ to the correlation function. For LCSRs, which rely on an
expansion in higher twist rather than higher condensates, we
correspondingly require
the contribution of higher twists to the LCSR  not to exceed
10\%. One  more requirement on the $s_0$ is that
they not stray away too much from ``reasonable'' values: $s_0$ is
to separate the ground state from higher mass contributions, and hence
should be below the next known clear resonance in that
channel. Assuming an excitation energy of $0.4$ to $0.8\,$GeV, we thus
expect the $s_0$ to lie in the interval 32 to 37~GeV$^2$, which is
evidently fulfilled by all $s_0$ quoted in Tab.~\ref{tab:7FF}.

\begin{table}
\addtolength{\arraycolsep}{3pt}\renewcommand{\arraystretch}{1.3}
$$
\begin{array}{|l|c|ccc||ccc|}
\hline & m_b & s_0 & M^2 & f_{B_q} &
 s_0 &  M^2 & f_{B_s}  \\\hline
{\rm set} 1 & 4.85 & 33.8 & 3.8 & 148 & 34.9 & 4.2 & 169 \\
{\rm set} 2 & 4.80 & 34.2 & 4.1 & 161 & 35.4 & 4.4 & 183 \\
{\rm set} 3 & 4.75 & 34.6 & 4.4 & 174 & 35.9 & 4.6 & 197 \\
\hline
\end{array}
$$
\addtolength{\arraycolsep}{-3pt}\renewcommand{\arraystretch}{1}
\caption[]{Parameter sets for $f_{B_q}$ and $f_{B_s}$ to $O(\alpha_s)$
  accuracy. $f_{B_q}$ and $f_{B_s}$
  are given in MeV, $s_0$ and $M^2$ in GeV$^2$. Note that
  the values of $f_{B_{q,s}}$ given in the table are
  {\it not} to be interpreted as 
meaningful determinations of these quantities,
  cf.\ text.}\label{tab:fB}
\end{table} 

Applying the above criteria to (\ref{eq:aa}), we obtain the sets of
$(s_0,M^2)$ collected in Tab.~\ref{tab:fB}, together with the
resulting $f_{B_{q,s}}$. We would like to stress
that these values are not to be interpreted as new independent
determinations of $f_{B_{q,s}}$, but are intermediate results to be used
in the evaluation of the LCSRs.

We proceed to determine the continuum thresholds and Borel parameters
for the LCSRs, using the same criteria as above. In order to keep
the complexity of the calculation at a manageable level,  for each form factor
the corresponding set is determined only once, at $q^2=0$.
To avoid confusion between parameters entering
\eqref{eq:aa} and those entering the LCSRs, let us call the latter
ones $M_{\rm LC}^2$ and $s_0^F$ where $F$ is the form factor. 
For larger $q^2$, these
parameters are expected to change slightly. Part of this effect can
be taken into account in the following way:
the tree-level LCSR to twist-2 accuracy reads, basically,
$$
\int_{u_0}^{1} du \,\frac{\phi(u)}{u}\,
e^{-(m_b^2-(1-u)q^2)/(u M_{\rm LC}^2)} \quad\text{with}\quad
u_0=\frac{m_b^2-q^2}{s_0-q^2} \,,
$$
which implies that the expansion parameter is $uM_{\rm LC}^2$ rather
than $M_{\rm LC}^2$.
We hence rescale the Borel parameter as
$$M_{\rm LC}^2\to M_{\rm LC}^2/\bra u \ket(q^2)$$
with the average value of $u$, $\bra u \ket(q^2)$, given by
\begin{equation*}
\bra u \ket(q^2) \equiv 
\left(\int_{u_0}^{1} du \,u \, \frac{\phi_(u)}{u}\, 
e^{-(m_b^2-(1-u)q^2)/(u M_{\rm LC}^2)}\right) /\left(
\int_{u_0}^{1} du\,\frac{\phi(u)}{u}\, e^{-(m_b^2-(1-u)q^2)/(u
    M_{\rm LC}^2)}\right)
\end{equation*}
with, approximately,
$\bra u \ket (0\,\text{GeV}^2)=0.86$ and 
$\bra u \ket (14\,\text{GeV}^2)=0.77$. The optimum Borel parameter
hence becomes larger with increasing $q^2$, which agrees with what one
finds when  $M_{\rm LC}^2$ is determined without rescaling.
Pa\-ra\-me\-tri\-sing the relation
between the Borel parameters of local and light-cone correlation functions as
\begin{equation}\label{eq:borels}
M_{\rm LC}^2 \equiv c_c M^2/\bra u \ket,
\end{equation}
we obtain, for $B_q \to \rho$, the values collected in Tab.~\ref{tab:7FF}. 
The sets for other transitions are  similar.

\begin{table}
\addtolength{\arraycolsep}{3pt}\renewcommand{\arraystretch}{1.3}
$$
\begin{array}{|l|c|cc|cc|cc|cc|cc|cc|}
\hline
  & m_b & s^V_0 & c_c^V & s^{A_0}_0 & c_c^{A_0} & s^{A_1}_0 & c_c^{A_1}
 & s^{A_2}_0 & c_c^{A_2} & s^{T_1}_0 & c_c^{T_1} & s^{T_3}_0 &
  c_c^{T_3}
  \\ \hline
{\rm set~1} & 4.85 & 35.2 & 1.7 & 33.0 & 1.7 & 33.7 & 1.7 & 34.1 & 1.7 &
  34.8 & 
1.7 & 34.7 & 1.7 \\
{\rm set~2} & 4.80 & 35.8 & 2.1 & 33.6 & 1.6 & 34.2 & 1.8 & 34.7 & 1.8 &
  35.3 & 
1.9 & 35.2 & 1.8 \\
{\rm set~3} & 4.75 & 36.4 & 2.1 & 34.2 & 1.6 & 34.7 & 1.9 & 35.3 & 1.9 &
  35.8 & 
2.1 & 35.7 & 1.9 \\
\hline
\end{array}
$$
\addtolength{\arraycolsep}{-3pt}\renewcommand{\arraystretch}{1}
\caption[]{Parameter sets for $B_q \to \rho$ for 
$V$, $A_0$, $A_1$, $A_2$, $T_1$ and $T_3$. 
As $T_1(0)=T_2(0)$
the corresponding parameters are equal.
$s_0$ and $M^2$ in GeV$^2$.}\label{tab:7FF}
\end{table} 

\subsection{Models for Distribution Amplitudes}\label{sec:4.3}

As mentioned in Sec.~\ref{sec:2} and detailed in App.~\ref{app:A}, the
DAs entering the LCSRs can be modelled by a truncated conformal
expansion. It turns out that the dominant contributions to the sum
rules come from the twist-2 DAs $\phi_{\perp,\parallel}$, which to NLO
in the conformal expansion are described by the lowest three Gegenbauer moments:
$a_0^{\perp,\parallel}\equiv 1$, which follows from the normalisation
of the DAs, $a_1^{\perp,\parallel}$, which is nonzero only for $K^*$, and
$a_2^{\perp,\parallel}$. In Ref.~\cite{BB98}, it was these three
parameters 
that were used to define 
the models for 
$\phi_{\perp,\parallel}$; all terms
$a_{n\geq 3}$ were dropped.

The numerical values of $a_{1,2}$ (and higher moments) are largely unknown.
$a_1$ has been determined from QCD sum rules in
\cite{CZreport,SU(3)breaking,moreSU(3)}. Averaging over the
determinations, we choose
\begin{equation}\label{a1}
a_1^\parallel(K^*,1\,{\rm GeV}) = 0.10\pm 0.07 =
a_1^\perp(K^*,1\,{\rm GeV})
\end{equation}
as our preferred values. Note that positive $a_1$ 
refer to a $K^*$ containing an $s$ quark -- for
a $\bar K^*$ with an $\bar s$ quark, $a_1$ changes sign.

Predictions for $a_2^{\perp,\parallel}$ also come from QCD
sum rules \cite{wavefunctions,CZreport,SU(3)breaking,BB96} and read
\begin{equation}
\begin{array}[b]{l@{\quad}l@{\quad}l}
a_2^{\parallel}(\rho,1\,{\rm GeV}) = 0.18\pm 0.10, & 
a_2^{\parallel}(K^*,1\,{\rm GeV}) = 0.09\pm 0.05, & 
a_2^{\parallel}(\phi,1\,{\rm GeV}) = 0\pm 0.1,\\[5pt]
a_2^{\perp}(\rho,1\,{\rm GeV}) = 0.2\pm 0.1,\phantom{00} & 
a_2^{\perp}(K^*,1\,{\rm GeV}) = 0.13\pm 0.08, & 
a_2^{\perp}(\phi,1\,{\rm GeV}) = 0\pm 0.1.
\end{array}\label{41}
\end{equation}
All these determinations have to be taken {\em cum grano salis}, as
the sum rules do not exhibit a clear Borel-window and also
become increasingly
unreliable for larger $n$.\footnote{This is due to the different power
  behavior of perturbative and nonperturbative terms in $n$, cf.\
  Ref.~\cite{BB96}.} 

But even assuming $a_{1,2}$ were known to sufficient accuracy -- under
what conditions is a truncation of $\phi$ 
after $a_2$ is justified? We have seen in Sec.~\ref{sec:2} that after
the convolution with a smooth short-distance function $T$ the
contributions of higher $a_n$ fall off sharply. So the actual question
is not so much how the truncated expansion compares to the full
convolution integral, but rather how the neglected terms compare to
other terms, for instance originating from 3-particle DAs, which
are included in the LCSR. For instance, assuming 
$a_i\geq 0.05$ it is necessary to include $a_2^\parallel$ and
$a_{2,4,6,8}^\perp$ in order to match the size of the contributions
from quark-quark-gluon matrix elements, and even for $a_i\geq 0.01$
one still needs $a_2^\parallel$ and $a_{2,4}^\perp$. If, on the other
hand, one consistently neglects terms that contribute less than 1\% to
the form factor, one can drop nearly all contributions from
quark-quark-gluon matrix elements, unless their values as given in
App.~\ref{app:A} are grossly underestimated. If $a_i\geq 0.05$, one
then has to keep $a_{2,4,6}^\perp$, but can drop all
$a_{n>0}^\parallel$. The upshot is that, in view of the lack of
information on $a_n^{\perp,\parallel}$, it is a good idea to devise models for
$\phi_{\perp,\parallel}$ with a small number of parameters, possibly
tied to experimental observables, and a
well-defined ``tail'' of higher-order Gegenbauer moments. This task
is undertaken in Ref.~\cite{angi}.

\begin{figure}[tb]
$$\epsfxsize=0.5\textwidth\epsffile{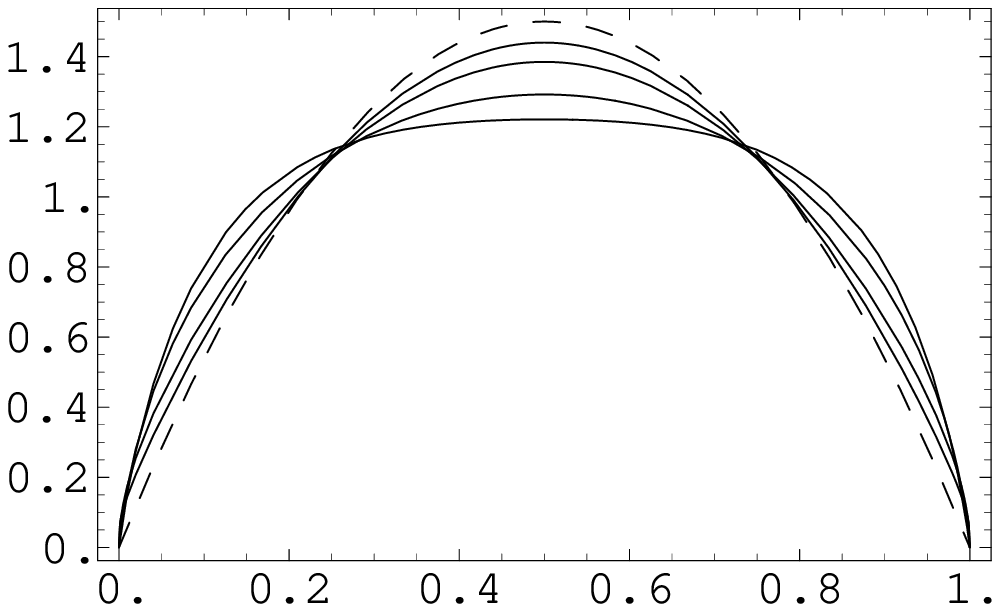}$$
\vskip-12pt
\caption[]{Examples for model DAs $\phi_a^+$ as functions of $u$, for
  $\Delta=1.2$ and 
$a=1.5,2,3,4$ (solid curves), as compared to the asymptotic DA
  (dashed curve). For $a\to 1$, $\phi^+_a$  approaches the asymptotic
  DA.}\label{fig:2}
$$\epsfxsize=0.5\textwidth\epsffile{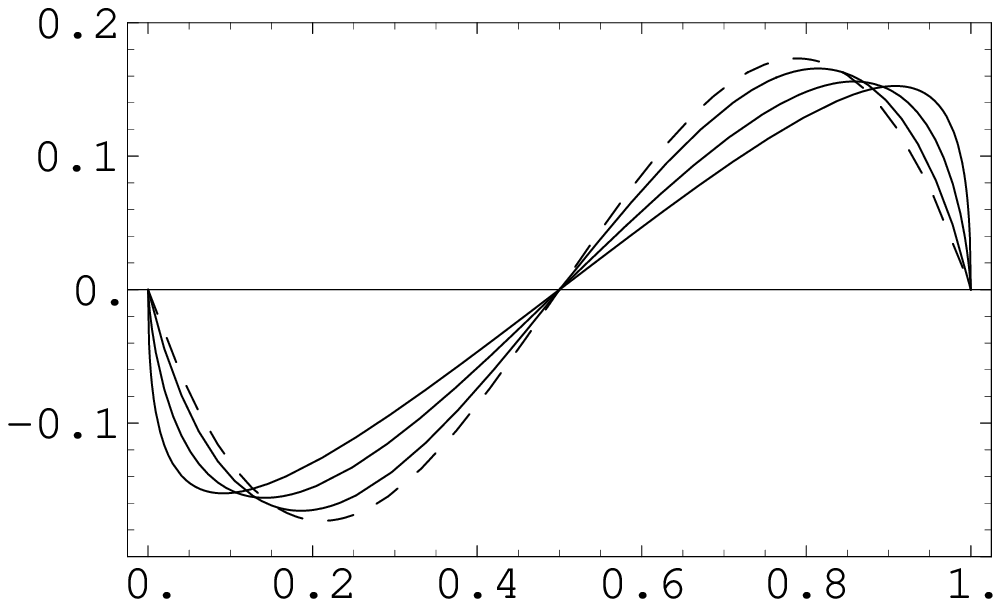}$$
\vskip-13pt
\caption[]{Models for the asymmetric contributions to the twist-2
  DA for $a_1 = 0.1$. Solid curves: $\phi_b^{\rm asym,+}$ 
as function of $u$ for
  $b\in\{2,3,5\}$; dashed curve: $\phi_\infty^{\rm asym,+}$. 
}\label{fig:3}
\end{figure}

Following Ref.~\cite{angi}, we introduce
two-parameter models for (the symmetric part of) 
$\phi_{\perp,\parallel}$ which are defined by the
fall-off behavior of the Gegenbauer moments 
$a_n$ in $n$ and the value of the integral
\begin{equation}\label{def:delta}
\Delta = \int_0^1 du \,\frac{\phi(u)}{3u} 
\equiv 1 + \sum_{n=1}^\infty a_{2n}\,;
\end{equation}
$\Delta=1$ for the asymptotic DA.
In particular we require $\Delta$ to be finite, which implies that the
$a_n$ must fall off sufficiently fast. The choice of $\Delta$
as characteristic parameter of $\phi$ relies on the fact that it is
directly related to an experimental observable, at least for $\pi$ and
$\eta$, namely the
$\pi(\eta)\gamma\gamma^*$ transition form factor, 
for which experimental constraints exist from CLEO \cite{CLEO}. 
We assume that the vector meson DAs are not fundamentally different
and take the range of $\Delta(\pi)$ extracted from CLEO as the likely
range for $\Delta(\rho)$. 
The second parameter characterizing our models is the
fall-off behavior of the Gegenbauer moments in $n$, which we assume to
be powerlike. We then can define a model DA $\tilde\phi_a^+$ in terms
of its Gegenbauer moments
\begin{equation}
\label{eq:anmodel}
a_n = \frac{1}{(n/2 + 1)^a}\,:
\end{equation}
using the generating function of the Gegenbauer-polynomials,
$$
f(\xi,t) = \frac{1}{(1-2 \xi t + t^2)^{3/2}} = \sum_{n=0}^\infty
C_n^{3/2}(\xi)\, t^n,
$$
this model can be summed to all orders in the Gegenbauer-expansion: 
\begin{eqnarray}
\tilde\phi_a^+(u) &=& \frac{3u \bar u }{\Gamma(a)}\, 
\int_0^1 dt (-\ln t)^{a-1}\, \left( f(2u-1,\sqrt{t})
+ f(2u-1,-\sqrt{t})\right).
\end{eqnarray}
The corresponding value of $\Delta$ is $\Delta_a^+ =
\zeta(a)$.
In order to obtain models for arbitrary
values of $\Delta$, we split off the asymptotic DA and write
\begin{eqnarray}
\phi^+_a(\Delta) &=& 6 u \bar u + \frac{\Delta-1}{\Delta_a^+ -1}
\left(\tilde{\phi}^+_a(u) - 6 u \bar u\right).\label{47}
\end{eqnarray}
Evidently one recovers the asymptotic DA for $\Delta=1$ and the
truncated conformal expansion with $a_2=\Delta-1$ and 
$a_{n\geq 4}=0$ for $a\to\infty$. 
The above formula is only valid for
$a>1$, as otherwise $\Delta^+_a$ diverges, or, equivalently,
$\phi^+_a$ does not vanish at the endpoints $u=0,1$. In
Fig.~\ref{fig:2} we plot several examples of $\phi_a^+$ for a fixed
value of $\Delta$. 

Our preferred values for $\Delta$, $a$ and the corresponding values of
$a_{2,4}^{\perp,\parallel}(1\,{\rm GeV})$ are collected in
Tab.~\ref{tab:Delta}. We choose $a=3\pm 1$ in order to obtain 
nonnegligible effects from higher order $a_n$. The choice of 
$\Delta(\rho)$ is motivated by the fact that all available
calculations indicate $a_2>0$, hence $\Delta>1$. We then fix the
maximum $\Delta$ in such a way that it yields $a_2<0.2$, which, given
the fact that the sum rule results (\ref{41}) are likely to overshoot
the true value of $a_2$, appears as to be the likely maximum value. We then
obtain $\Delta(\rho) = 1.15\pm 0.10$, with a rather conservative
error. We choose the same values for $\omega$. 
For $K^*$ and $\phi$, we take into account that the values of
$a_2$ appear to have the tendency to decrease, which was noticed
already in Ref.~\cite{CZreport}. Assuming that the decrease is 20\% from
$\rho$ to $K^*$, and another 20\% from $K^*$ to $\phi$, we arrive at
the numbers quoted in Tab.~\ref{tab:Delta}.
\begin{table}
\renewcommand{\arraystretch}{1.2}\addtolength{\arraycolsep}{3pt}
$$
\begin{array}{|l|cc|cc|}
\hline
& \Delta & a & a_2^{\perp,\parallel}(1\,{\rm GeV}) & 
a_4^{\perp,\parallel}(1\,{\rm GeV})\\\hline
\rho,\omega & 1.15\pm 0.10 & 3\pm 1 & 0.09^{+0.10}_{-0.07} & 0.03\pm0.02\\
K^* & 1.12\pm 0.10 & 3\pm 1 & 0.07^{+0.09}_{-0.07} & 0.02\pm 0.01\\
\phi & 1.10\pm 0.10 & 3\pm 1 & 0.06^{+0.09}_{-0.07} & 
0.02^{+0.01}_{-0.02}\\\hline
\end{array}
$$
\renewcommand{\arraystretch}{1}\addtolength{\arraycolsep}{-3pt}
\caption[]{Values  for $\Delta(1\,{\rm
    GeV})\equiv\Delta^{\perp,\parallel}(1\,{\rm GeV})$, $a$ and
the corresponding values of 
$a_{2,4}^{\perp,\parallel}(1\,{\rm GeV})$.}\label{tab:Delta}
\end{table}

Evidently, DAs defined in dependence of $\Delta$ also require a
specification of the scale at which they are valid. As we presume that
$\Delta(\mu)$ will be measured, if at all, at the low scale $\mu=1\,{\rm
  GeV}$, we choose this as the reference scale. $\Delta$ at higher
scales can be obtained from Eq.~(\ref{def:delta}), using the leading-order
RG-improved expressions for $a_n(\mu)$, or, if $\mu$ is not too
different from 1~GeV, from the unimproved expression Eq.~(\ref{25}).

Models for the asymmetric part of the DA, relevant for $K^*$, can be
constructed in a similar way as
\begin{eqnarray}
\tilde\psi_b^+(u) &= &\frac{3u \bar u }{\Gamma(b)}\, 
\int_0^1 dt (-\ln t)^{b-1}\, \left( f(2u-1,\sqrt{t})
- f(2u-1,-\sqrt{t})\right).
\end{eqnarray}
One relevant parameter is $b$, and as the second one we choose $a_1$. 
Models for the asymmetric part of $\phi$ with arbitrary $a_1$ can then
be defined as
\begin{equation}
\phi_b^{{\rm asym},+} = a_1 (3/2)^b \tilde\psi_b^+(u).
\end{equation}
Examples for such models are shown in Fig.~\ref{fig:3}.

$\phi_b^{\rm asym,+}$ also contributes to the value of $\Delta$:
\begin{eqnarray*}
\Delta^{{\rm asym},+} & = & \int_0^1 du\,\frac{\phi_b^{\rm
    asym,+}(u)}{3u} = - a_1 (3/2)^b
\zeta(b,3/2),
\end{eqnarray*}
where $\zeta(b,s)= \sum_{k=0}^\infty 1/(k+s)^b$ is the Hurwitz $\zeta$
function.
Our models for the $K^*$ DA are hence characterized by four parameters:
$\Delta$, $a$ of the symmetric part and $a_1$, $b$ of the asymmetric
part. The total value of $\Delta$ is given by 
$$\Delta^{{\rm total},+}  = \Delta + \Delta^{{\rm asym},+}.$$
In the actual calculation we choose $a=b$.

\subsection{\boldmath Results for $q^2=0$}\label{sec:4.4}

\begin{table}[tbp]
\renewcommand{\arraystretch}{1.1}\addtolength{\arraycolsep}{5pt}
$$
\begin{array}{|c|c||cccc||c|c|}
\hline
& F(0) & \Delta_{m_b} & \Delta_{7p} &
\Delta_L & \Delta_T & \Delta_{\rm tot} & \Delta_{a_1}  \\ \hline
V^{B_q \to \rho} & 0.323 & 0.007& 0.025 & 0.005 & 0.013 & 0.029 &  \\
A_0^{B_q \to \rho} & 0.303 & 0.004 & 0.026 & 0.009 & 0.006 & 0.028 &  \\
A_1^{B_q \to \rho} & 0.242 & 0.007 & 0.020 & 0.004 & 0.010 & 0.024 &  \\
A_2^{B_q \to \rho} & 0.221 & 0.008 & 0.018 & 0.002 & 0.011 & 0.023 &  \\
T_1^{B_q \to \rho} & 0.267 & 0.004 & 0.018 & 0.004 & 0.010 & 0.021 &  \\
T_3^{B_q \to \rho} & 0.176 & 0.001 & 0.013 & 0.001 & 0.009 & 0.016 &  \\ 
\hline
V^{B_s \to \rar{K}^*} & 0.311 & 0.006 & 0.021 & 0.003 & 0.013 & 0.026 & 
-0.43\,\relta_{a_1} \\
A_0^{B_s \to \rar{K}^*}  & 0.363 & 0.003 & 0.032 & 0.006 & 0.009 & 0.034 & 
-0.37\,\relta_{a_1} \\
A_1^{B_s \to \rar{K}^*} & 0.233 & 0.007 & 0.019 & 0.002 & 0.010 & 0.023 & 
-0.32\,\relta_{a_1} \\
A_2^{B_s \to \rar{K}^*} & 0.181 & 0.008 & 0.021 & 0.001 & 0.010 & 0.025 & 
-0.30\,\relta_{a_1} \\
T_1^{B_s \to \rar{K}^*} & 0.260 & 0.005 & 0.021 & 0.003 & 0.010 & 0.024 & 
-0.33\,\relta_{a_1} \\
T_3^{B_s \to \rar{K}^*} & 0.136 & 0.003 & 0.013 & 0.000 & 0.008 & 0.016 & 
-0.17\,\relta_{a_1} \\ 
\hline
V^{B_q \to K^*} & 0.411 & 0.008 & 0.029 & 0.003 & 0.013 & 0.033 & 
\phantom{-}0.44\,\relta_{a_1} \\
A_0^{B_q \to K^*} & 0.374 & 0.009 & 0.031 & 0.005 & 0.008 & 0.034 & 
\phantom{-}0.39\,\relta_{a_1} \\
A_1^{B_q \to K^*} & 0.292 & 0.009 & 0.025 & 0.002 & 0.009 & 0.028 & 
\phantom{-}0.33\,\relta_{a_1} \\
A_2^{B_q \to K^*} & 0.259 & 0.009 & 0.023 & 0.001 & 0.010 & 0.027 & 
\phantom{-}0.31\,\relta_{a_1} \\
T_1^{B_q \to K^*} & 0.333 & 0.005 & 0.026 & 0.003 & 0.010 & 0.028 & 
\phantom{-}0.34\,\relta_{a_1} \\
T_3^{B_q \to K^*} & 0.202 & 0.002 & 0.016 & 0.001 & 0.008 & 0.018 & 
\phantom{-}0.18\,\relta_{a_1} \\
\hline
V^{B_q \to \rmega} & 0.293 & 0.006 & 0.025 & 0.002 & 0.013 & 0.029 &  \\
A_0^{B_q \to \rmega} & 0.281 & 0.012 & 0.027 & 0.003 & 0.006 & 0.030 &  \\
A_1^{B_q \to \rmega} & 0.219 & 0.008 & 0.021 & 0.001 & 0.010 & 0.025 &  \\
A_2^{B_q \to \rmega} & 0.198 & 0.007 & 0.018 & 0.001 & 0.011 & 0.022 &  \\
T_1^{B_q \to \rmega} & 0.242 & 0.003 & 0.019 & 0.002 & 0.010 & 0.022 &  \\
T_3^{B_q \to \rmega} & 0.155 & 0.000 & 0.012 & 0.000 & 0.009 & 0.015 &\\ 
\hline
V^{B_s \to \rhi} & 0.434 & 0.004 & 0.032 & 0.003 & 0.014 & 0.035 &  \\
A_0^{B_s \to \rhi} & 0.474 & 0.002 & 0.031 & 0.005 & 0.019 & 0.037 &  \\
A_1^{B_s \to \rhi} & 0.311 & 0.007 & 0.027 & 0.002 & 0.009 & 0.029 &  \\
A_2^{B_s \to \rhi} & 0.234 & 0.011 & 0.024 & 0.001 & 0.009 & 0.028 &  \\
T_1^{B_s \to \rhi} & 0.349 & 0.004 & 0.031 & 0.002 & 0.010 & 0.033 &  \\
T_3^{B_s \to \rhi} & 0.175 & 0.003 & 0.016 & 0.000 & 0.007 & 0.018 &  \\
\hline
\end{array}
$$
\vskip-5pt
\renewcommand{\arraystretch}{1}\addtolength{\arraycolsep}{-5pt}
\caption[]{Form 
factors at $q^2=0$ for parameter set 2 of Tabs.~\ref{tab:fB}
and \ref{tab:7FF}, i.e.\ $m_b=4.8\,$GeV. The form factors are defined
in Eqs.~(\ref{eq:SLFF}) and (\ref{eq:pengFF}). The penguin form factors
$T_i$ are evaluated at the UV scale $\mu = m_b$.
$\Delta_{m_b}$ is the variation of the result with $m_b$, i.e.\ the
maximum deviation between the results obtained for sets 1, 2 and 3.
$\Delta_{7p}$ is the maximum deviation found by scanning the
7-parameter space discussed in the text. $\Delta_L$
and $\Delta_T$ are the uncertainties induced by the vector and tensor 
couplings in
Tab.~\ref{tab:fV}. The total error $\Delta_{\rm tot}$ is obtained by adding
$\Delta_{(m_b,7p,L,T)}$ in quadrature. Form factors involving $K^*$
carry one more uncertainty $\Delta_{a_1}$ induced by the Gegenbauer moment
$a_1$, with $\delta_{a_1} = [a_1({K^*},1\,{\rm GeV})-0.1]$. 
}\label{tab:basic}
\end{table} 

Let us first analyse the form factors for $q^2=0$.
Using the input parameters given in Tabs.~\ref{tab:7FF} and
\ref{tab:Delta}, we obtain the results collected in
Tab.~\ref{tab:basic}. 

\begin{figure}[tb]
$$\begin{array}{@{}c@{\quad}c@{}}
\epsfxsize=0.45\textwidth\epsffile{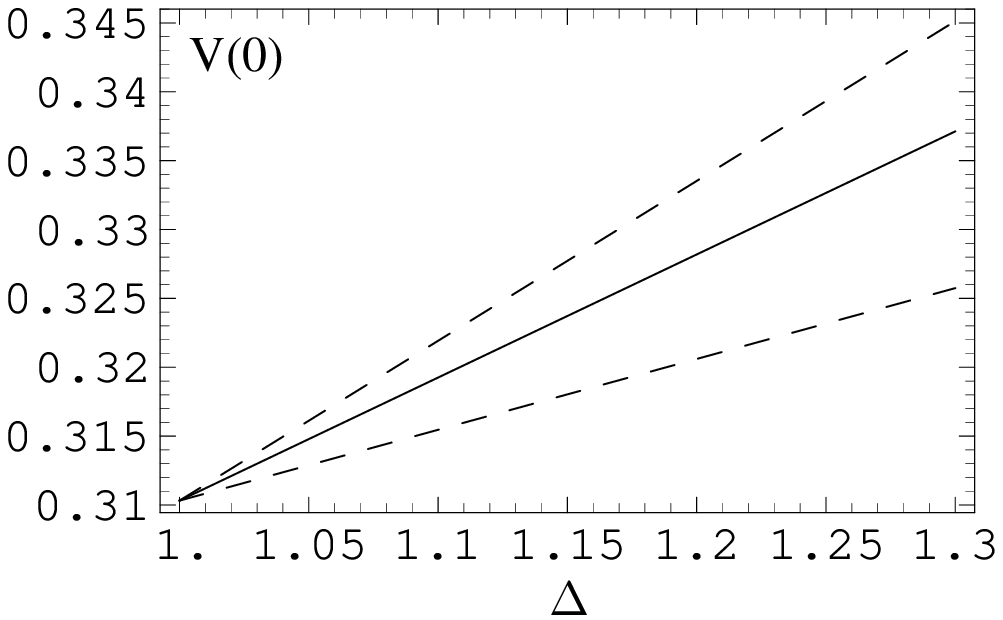} &
\epsfxsize=0.45\textwidth\epsffile{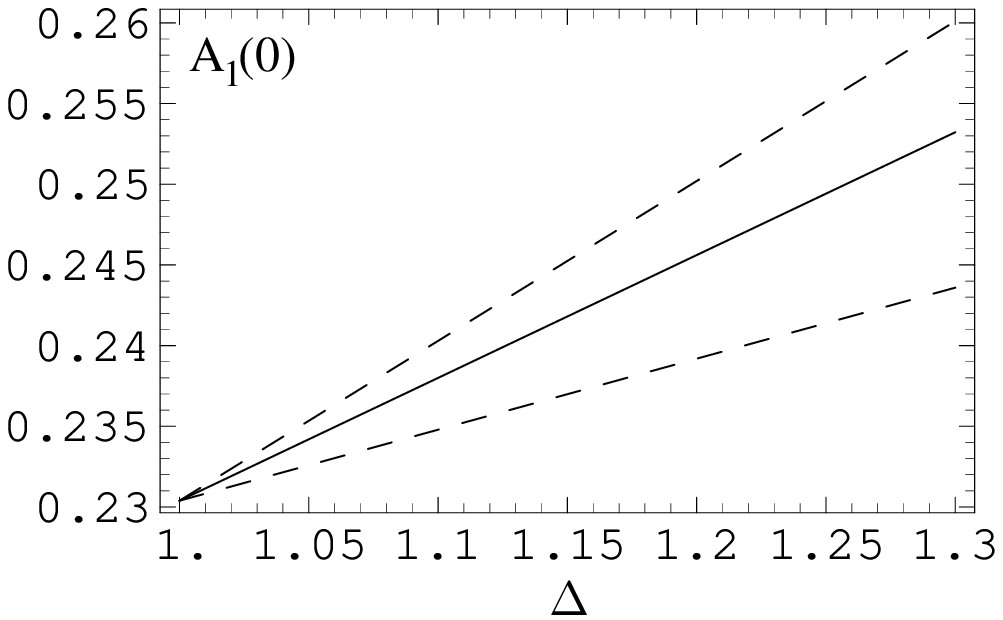}\\[10pt]
\epsfxsize=0.45\textwidth\epsffile{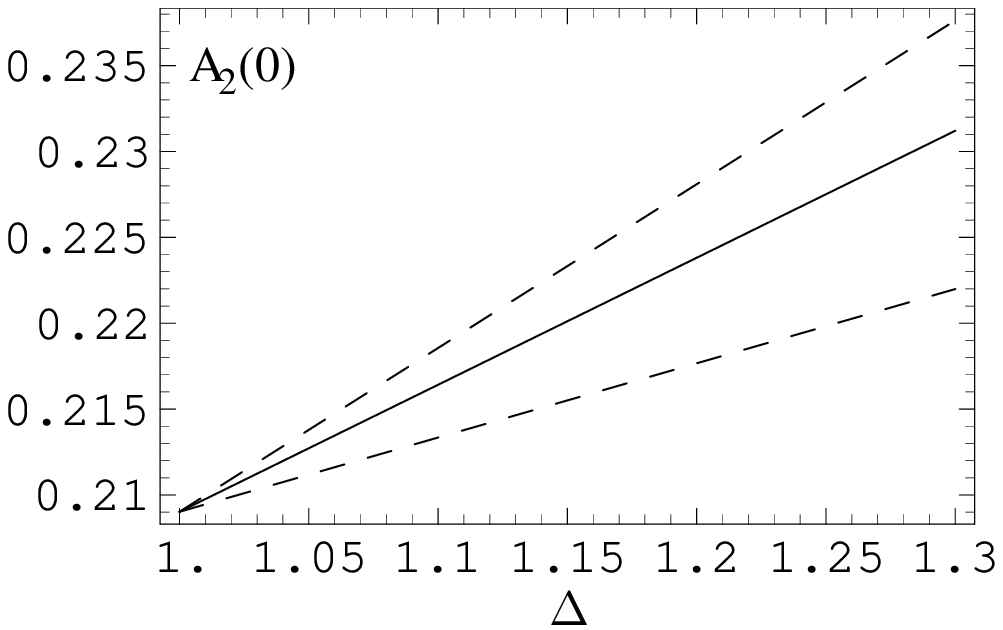} &
\epsfxsize=0.45\textwidth\epsffile{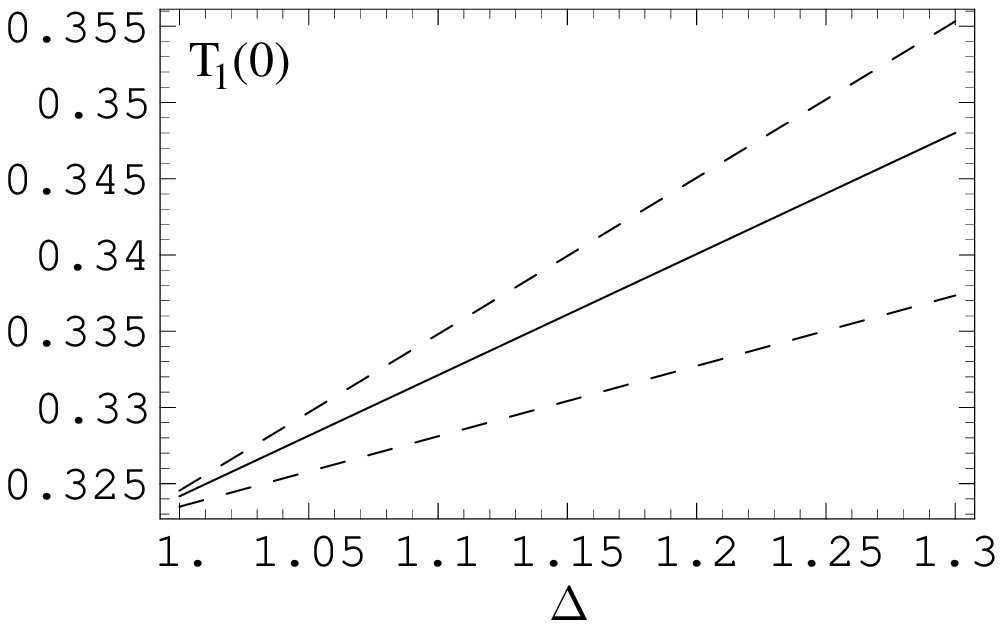}\\[-0.6cm]
\end{array}
$$
\caption[]{The form factors $V(0)$, $A_1(0)$, $A_2(0)$ for
  $B_q\to\rho$ and $T_1(0)$ for $B_q\to K^*$ as functions of $\Delta$,
  the main parameter of the twist-2 DAs. Solid lines: central values of
  input parameters. Dashed lines: variation of the form factors with a
  change of $a$, the second parameter of the DAs, by $\pm 1$. Allowed
  values of $\Delta$: cf.\ Tab.~\ref{tab:Delta}.}\label{fig:del}
\end{figure}

For the discussion of theoretical uncertainties, we distinguish
between uncertainties that can be reduced by future more accurate
determinations of the corresponding hadronic parameters and others
that are either systematic uncertainties, inherent to the method of
LCSRs, or parameter uncertainties 
not likely to be reduced in the near future.
The latter comprise the dependence of the form factors on the LCSR
parameters $s_0$, $M^2$, $\mu_{\rm IR}$ and, via $f_B$, the quark and
mixed condensate. Our results also depend, very mildly, on $m_s$ and,
more importantly, on the meson DAs which are described by the
2-parameter model (\ref{47}). All these parameters induce a
theoretical error of
the form factors which we determine by varying 
\begin{itemize}
\item the threshold $s_0$ by $\pm 1.0\,\text{GeV}^2$;
\item the Borel parameter
$M^2$ in Eq.~\eqref{eq:borels} by $\pm 1.5\,\text{GeV}^2$;
\item the infrared factorization scale $\mu_{\rm IR}=\sqrt{m_B^2-m_b^2}$ 
by $\pm 1\,\text{GeV}$;
\item the quark condensate and the mixed condensate as indicated in 
Eq.~\eqref{eq:conds};
\item the first inverse moment of the twist-2 DAs, $\Delta$, by $\pm 0.1$;
\item the power behavior of the Gegenbauer moments, $a$, by $\pm 1$;
\item the strange quark mass $m_s$ by $\pm 20 \%$. 
\end{itemize}
The largest deviation of the form factor from its central value, 
in this 7-parameter space, is dubbed $\Delta_{7p}$ and amounts 
to typically 7 to 11\%. In Fig.~\ref{fig:del} we show the dependence
of selected form factors on $\Delta$ and $a$. The uncertainty in these
parameters is the most important single source of error of the form
factors and amounts to half of the total error.

The form factors also depend, rather mildly, on $m_b$: varying $m_b$
by $\pm 0.05\,$GeV around the central value 4.8$\,$GeV, and using
$s_0$ and $M^2$ as given in Tabs.~\ref{tab:fB} and \ref{tab:7FF}, we
obtain the error $\Delta_{m_b}$ which ranges from 1\% to 5\%.

One more source of uncertainty of the form factors is due to
$f_V$ and $f_V^T$, the vector and tensor coupling of the vector
mesons. This is easily
understood by splitting the generic form factor $F$ into two terms 
proportional to $f^T$ and
$f^L\equiv f$:
\begin{equation}
F = f^L F^L + f^T F^T.
\end{equation}
As argued in Sec.~\ref{sec:2}, the first term is of order
$\delta\propto m_V$, the second of order 1 and indeed, for most form
factors, is the dominant
contribution. The present errors of $f^T$, as collected in
Tab.~\ref{tab:fV}, are nonnegligible. 
$f^T$ is accessible to lattice calculations and
first results have been reported in Ref.~\cite{lattfT}, which
indicates that a reduction of
the error of $f^T$ seems feasible. In order to allow the
adjustment of our form factors to new results for $f^T$, we
give explicit results for $F^L$ and $F^T$ in
App.~\ref{app:0}. The uncertainties $\Delta_{T,L}$ 
of the form factors due to the present values of
$f^{T,L}$ are included in
Tab.~\ref{tab:basic}. $\Delta_T$ is typically of order 4\%, 
$\Delta_L$ is much smaller. 

For transitions involving the $K^*$, an additional uncertainty is
induced by the first Gegenbauer moment $a_1$, and is given by
$\Delta_{a_1}$ in Tab.~\ref{tab:basic}, where the quantity 
$\delta_{a_1}$ is defined as $[a_1({K^*},1\,{\rm GeV})-0.1]$. Note
that $a_1({K^*})$ refers to a $s\bar q$ bound state and hence 
$a_1({\bar K^*}) = - a_1(K^*)$, which explains the negative sign of
the corresponding entries in Tab.~\ref{tab:basic}. Again we aim to
make our results adjustable to any future improvement in the
determination of $a_1$ and
give explicit results for the corresponding contributions in 
App.~\ref{app:0}.

Some important features of the results collected in
Tab.~\ref{tab:basic} are:
\begin{itemize}
\item the form factors for $B_q\to K^*$ transition are about 20\%
  larger than those for $B_q\to\rho$. The reason for this is
  twofold: on the one hand, the $K^*$ vector and tensor couplings are
  larger than those of the $\rho$. On the other hand, the
  SU(3)-breaking of the twist-2 DAs, parametrised by the first
  Gegenbauer moment $a_1$, gives a positive contribution to the form
  factors;
\item the form factors for $B_s\to  K^*$ have a tendency to be
  smaller than those for $B_q\to\rho$. The reason for this is a
  negative contribution of $a_1$ and the fact that $f_{B_s}$ is larger
  than $f_{B_q}$. On the other hand, the optimum $s_0$ are also larger
  than for  $B_q\to\rho$, which partially compensates the first two effects;
\item the $B_q\to\omega$ form factors are slightly smaller than those
  for $B_q\to\rho$. This is a consequence of the fact that the
  $\omega$ vector and tensor couplings are smaller than those of the $\rho$;
\item the total theoretical error is dominated by that of the twist-2
  DAs and the sum rule parameters $s_0$ and $M^2$. The former  can,
  in principle, be reduced by future calculations, the second is
  systematic and irreducible.
\end{itemize}

The typical total uncertainty of each form factor is 10\%, 
ranging between 8\% and
13\%. Any significant reduction of the error requires more
accurate information on the twist-2 DAs. The minimum irreducible
theoretical uncertainty is set by the systematic uncertainty of the
LCSR approach and encoded in the dependence of the results on $s_0$
and $M^2$; it amounts to about 6 to 7\%.

Let us also compare our results to those obtained in Ref.~\cite{BB98} 
by the same method, but with less sophistication. The main difference
between our present and our previous analysis is the inclusion of
radiative corrections to 2-particle twist-3 contributions, the
description of the twist-2 DAs by models including all-order effects
in the conformal expansion,
the more accurate determination of the sum rule parameters $s_0$ and
$M^2$ and the much more detailed error analysis. The most striking
difference between the actual results affects the $B_s\to  K^*$
transition, whose form factors were predicted, in \cite{BB98}, 
to be between 10 and 30\% smaller than those in
Tab.~\ref{tab:basic}. The reason for this discrepancy is mainly 
the more accurate determination of $M^2$ and $s_0$ we
employ in the present analysis --- in Ref.~\cite{BB98} all
form factors were determined for the {\em same} values of $M^2$ and
$s_0$. All other form factors quoted in \cite{BB98} agree, within
$\pm 15$\%, with those of Tab.~\ref{tab:basic}, which is within the
theoretical uncertainty stated in \cite{BB98}. The only exception is
$T_3(0)$, which deviates by between 15 and 45\% from the numbers
obtained in \cite{BB98}. The reason for this discrepancy lies in the
(correct) treatment of factors $1/(pz)$ in our present paper, 
cf.\ Sec.~\ref{sec:3}.

\begin{figure}[tb]
$$
\begin{array}{@{}c@{\quad}c}
\epsfxsize=0.45\textwidth\epsffile{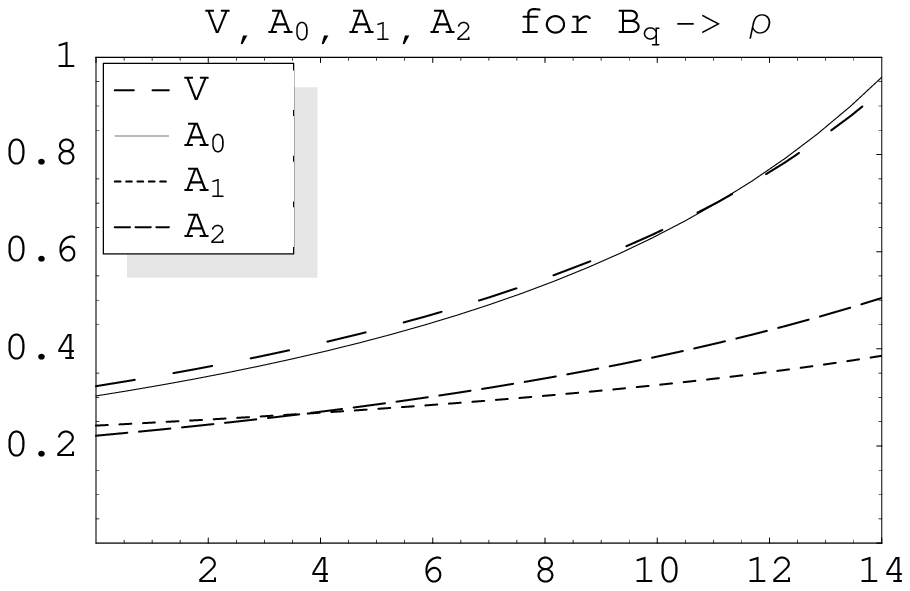} & 
\epsfxsize=0.45\textwidth\epsffile{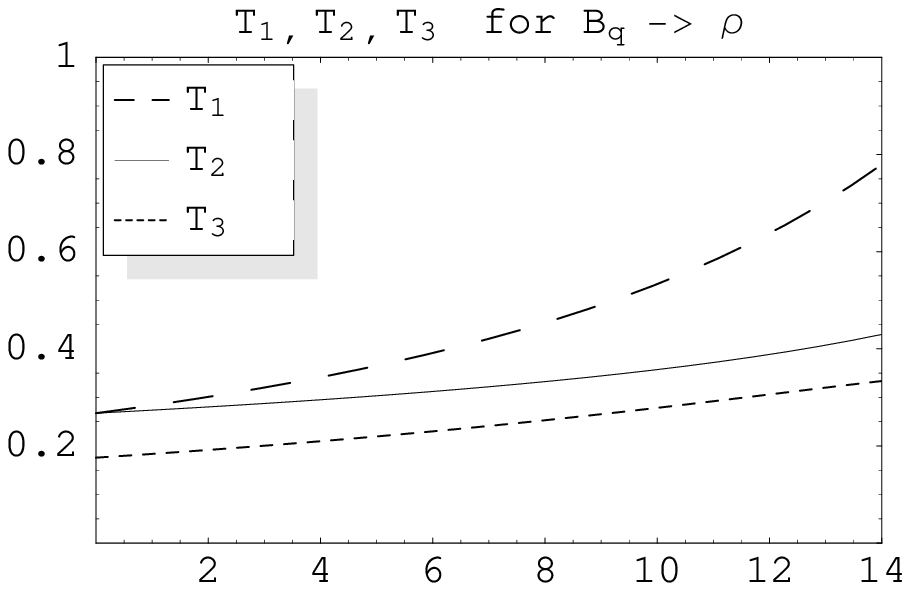}\\[5pt]
\epsfxsize=0.45\textwidth\epsffile{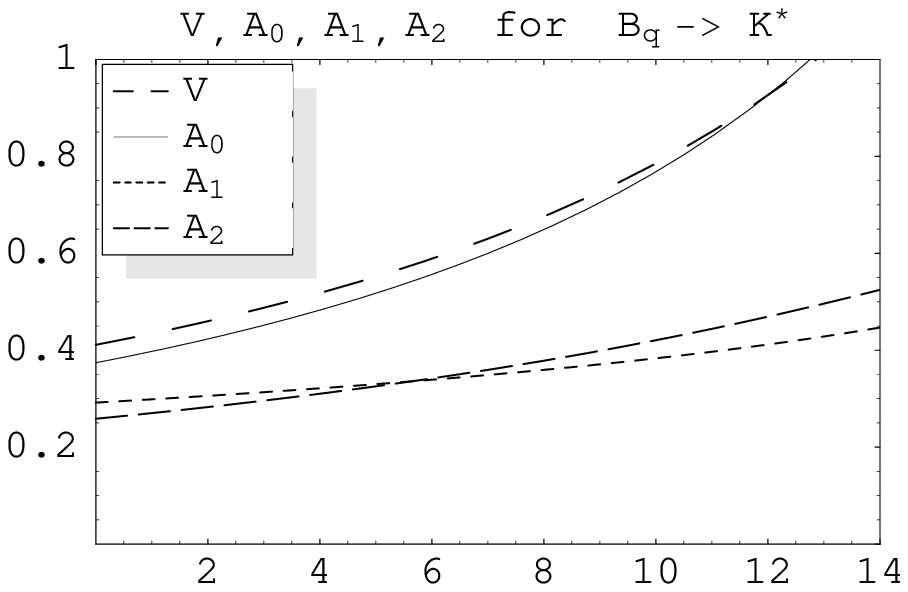} & 
\epsfxsize=0.45\textwidth\epsffile{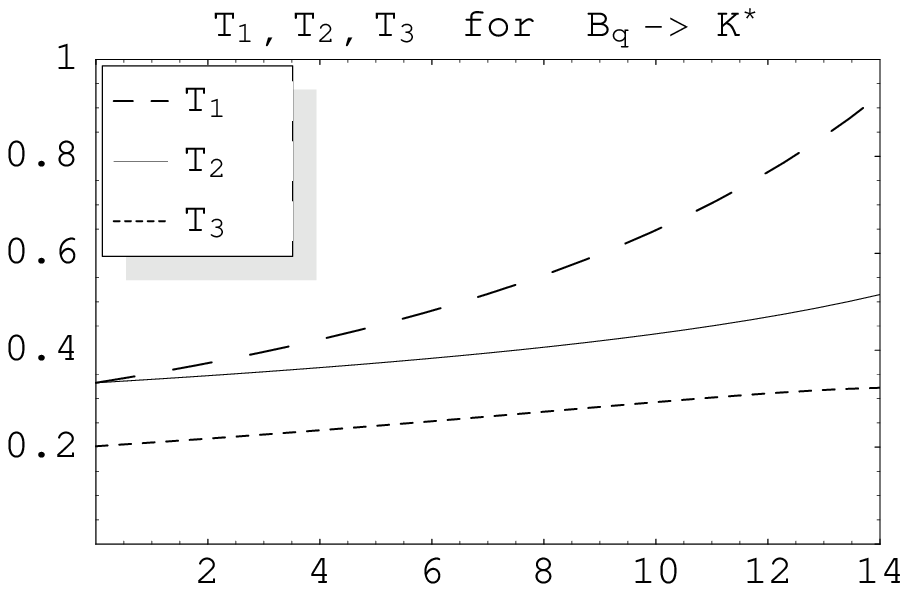}\\[5pt]
\epsfxsize=0.45\textwidth\epsffile{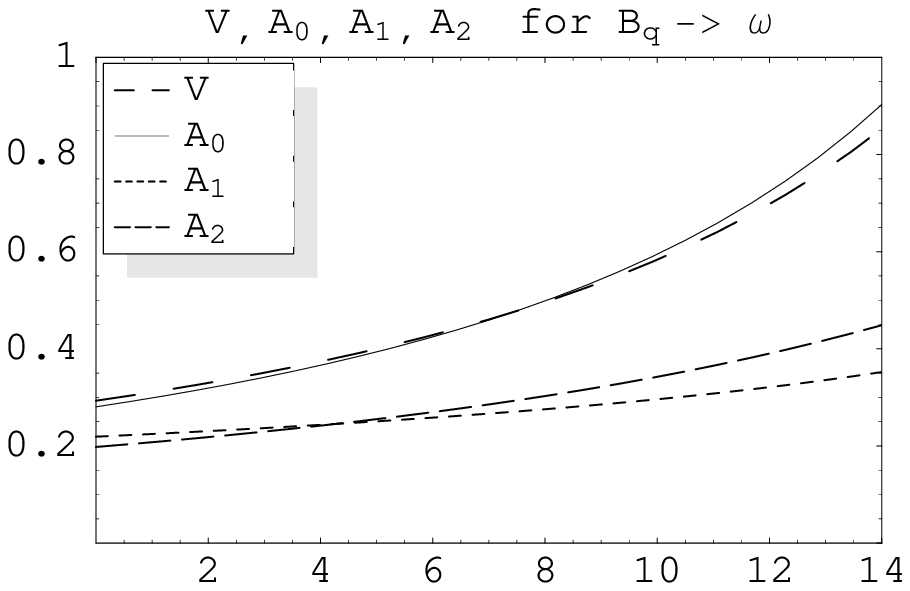} & 
\epsfxsize=0.45\textwidth\epsffile{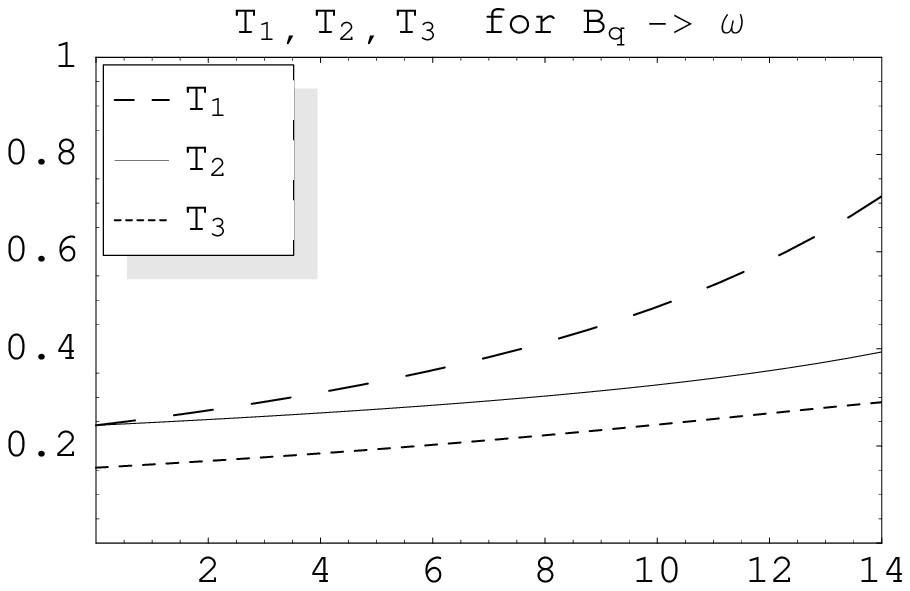}
\end{array}
$$
\vskip-10pt
\caption[]{Form factors for $B_q$ decays as functions of $q^2$,
for central values of input parameters.
}\label{fig:mainA}
\end{figure}

\begin{figure}[tb]
$$
\begin{array}{@{}c@{\quad}c}
\epsfxsize=0.45\textwidth\epsffile{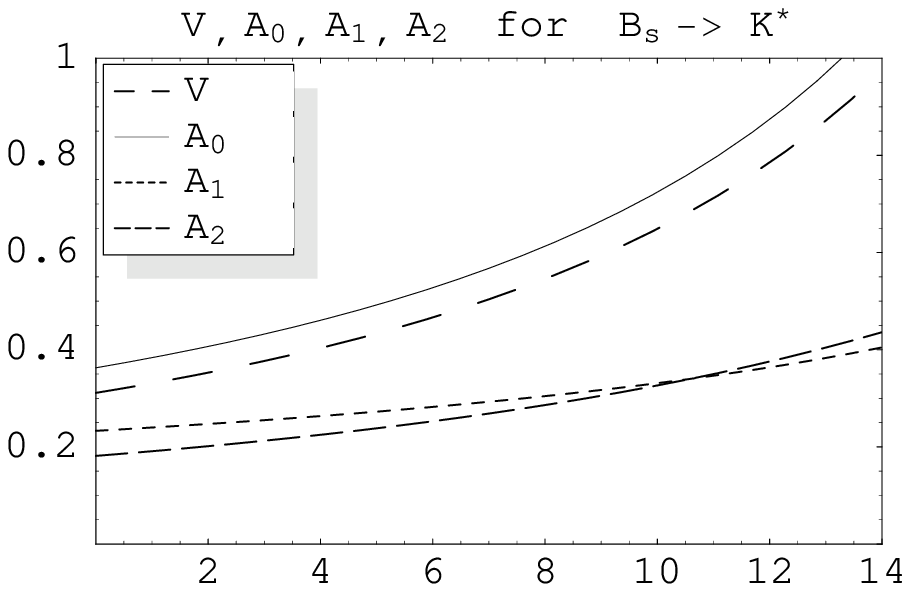} & 
\epsfxsize=0.45\textwidth\epsffile{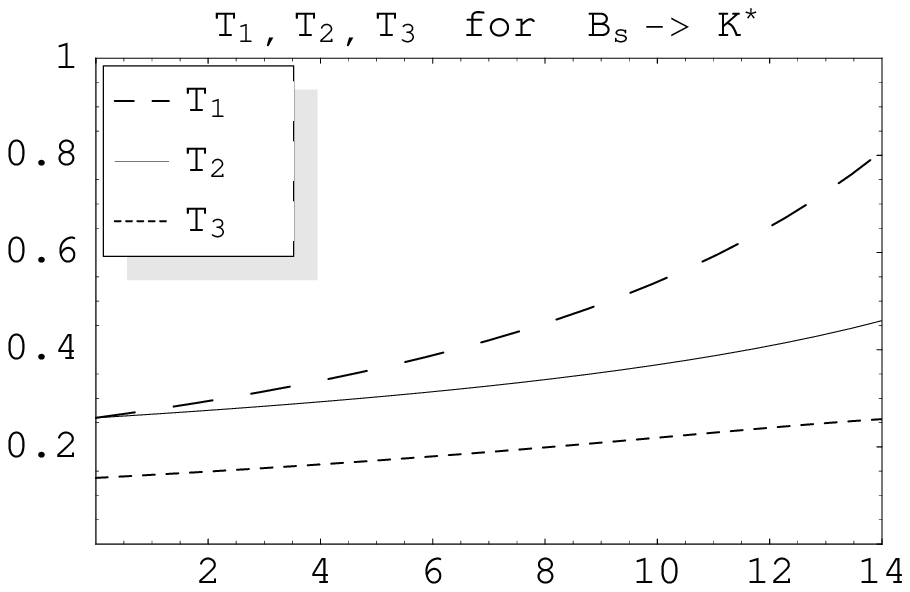}\\[5pt]
\epsfxsize=0.43\textwidth\epsffile{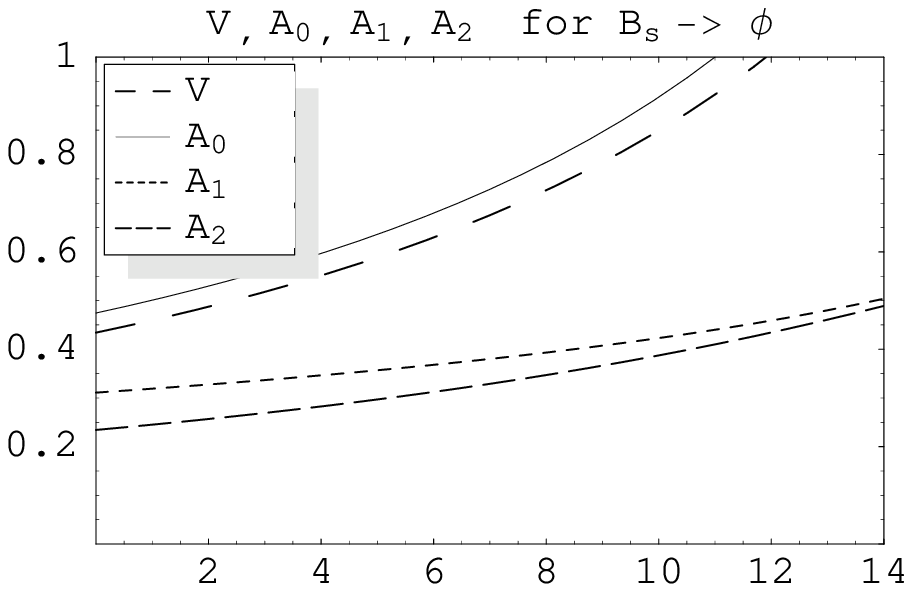} & 
\epsfxsize=0.43\textwidth\epsffile{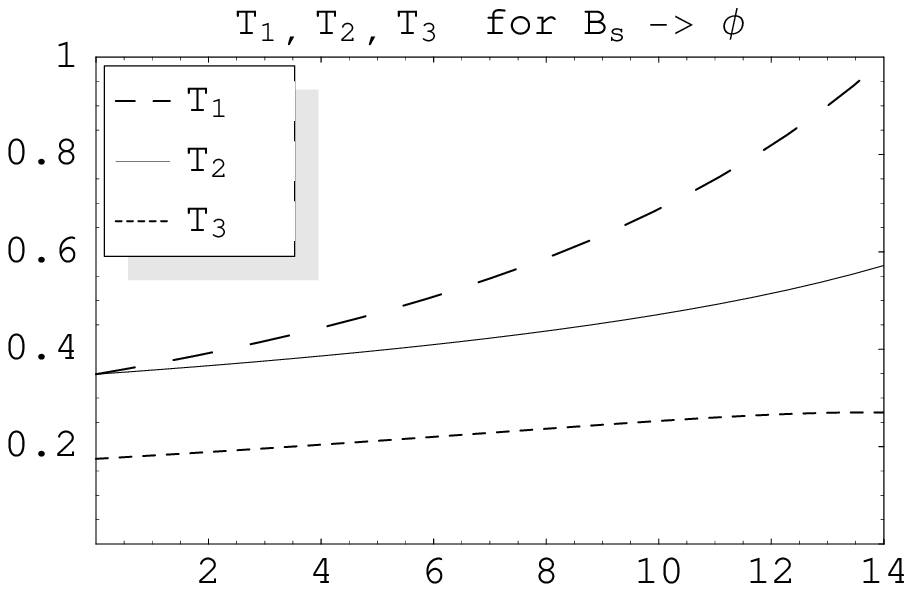}
\end{array}
$$
\vskip-10pt
\caption[]{Form factors for $B_s$ decays as functions of $q^2$,
for central values of input parameters.}\label{fig:mainB}
\end{figure}

\subsection{Results for $q^2\neq0$, Fits and Extrapolations}\label{sec:4.5}

In this subsection we discuss the $q^2$-dependence of the form factors. 
The results of the LCSR calculation are plotted in
Figs.~\ref{fig:mainA} and \ref{fig:mainB}.
They can be parametrised in terms of simple formulas with 2 or 3 parameters,
which are valid in the full kinematical regime in $q^2$. The
corresponding parameters are collected in Tab.~\ref{tab:fits}.

As mentioned in Sec.~\ref{sec:2}, the LCSR method is valid for large
energies of the final state vector meson, $E_V \gg \Lambda_{\rm QCD}$, 
which implies, via the relation $q^2 = m_B^2 - 2 m_B E_V$,
a restriction to not too large $q^2$.
We include values in the regime 
\begin{equation}
\label{eq:LCSRregime}
0\leq q^2\leq q^2_{\rm LCSR, max} = 14\,{\rm GeV}^2\,,
\end{equation}
which has to be compared with the maximum physical $q^2_{\rm phys, max} =
(m_B-m_V)^2$ of $20.3\,{\rm GeV}^2$ for $B_q\to(\rho,\omega)$, 
$19.2\,{\rm GeV}^2$ for $B_q \to K^*$, $20.0 \,{\rm GeV}^2$ for 
$B_s \to \bar{K}^*$ and $ 18.2\,{\rm GeV}^2$ for $B_s \to \phi$. The
main aim of this subsection is to provide fits for the LCSR
results, which are valid in the full physical regime of $q^2$. We
will comment below on the dependence of the fit results on the actual
value used for $q^2_{\rm LCSR, max}$.

We closely follow the procedure we used in our previous paper
on $B\to\,$pseu\-do\-sca\-lar form factors, Ref.~\cite{BZ}. Generically,
barring the occurence of anomalous thresholds, 
any form factor $F(q^2)$ has singularities (poles and cuts) 
for positive real $q^2$, starting at
the position of the lightest resonance coupling to the relevant current,
and hence can be written as a dispersion integral in $q^2$. 
Splitting off the lowest-lying resonance with mass $m_R$, one has
\begin{equation}
F(q^2)  = \frac{r_1}{1-q^2/m_R^2}+ 
\int_{t_0}^\infty
ds\,\frac{\rho(s)}{s-q^2}\,,\label{eq:xyz}
\end{equation}
where $t_0$ is the threshold for multiparticle contributions, which
can be above or below $m_R^2$.
Keeping only the first term and neglecting the integral altogether one
obtains the vector meson dominance (VMD)
approximation.\footnote{This notion comes from the analysis of
  electromagnetic form factors, where the first resonance is the
  $\rho$. In weak decays, however, the lowest resonance is, in general, not a
  vector meson, so that the notion VMD is, strictly speaking, obsolete.} 
Even though this approximation is
expected to work very well close to the pole, 
it certainly won't work far away from
it, e.g.\ at $q^2=0$. For $B\to \pi$ transitions it was
argued in Ref.~\cite{BecKai} that the integral can be modelled by a 
second pole at larger $q^2$, which is unrelated to any physical
resonance:
\begin{equation}\label{eq:dp}
F(q^2)=\frac{r_1}{1-q^2/m_R^2}+\frac{r_2}{1-q^2/m_{\rm fit}^2}
\end{equation}
with the three independent parameters $r_{1,2}$ and $m_{\rm fit}$.
\begin{table}[p]
\renewcommand{\arraystretch}{1}
\addtolength{\arraycolsep}{3pt}
\vskip-10pt
$$
\begin{array}{|l||c|cc||cccc|c|c|}
\hline
& F(0) & \Delta_{\rm tot} & \Delta_{a_1}   & r_1 & m_R^2 & r_2 &
m_{\rm fit}^2 & \delta  & \mbox{fit eq.} \\
\hline
V^{B_q \to \rho} & 0.323 & 0.030   & & \phantom{-}1.045 & m^2_{1^-} &
-0.721 & 38.34 & 0.1 & (\ref{eq:dp})\\ 
A_0^{B_q \to \rho} & 0.303 & 0.029 & & \phantom{-}1.527 & m^2_{0^-} & -1.220  & 33.36 & 0.1 &(\ref{eq:dp})\\ 
A_1^{B_q \to \rho} & 0.242 & 0.023 & & -     & -         & \phantom{-}0.240 & 37.51 & 1.0 &(\ref{eq:vmd})\\ 
A_2^{B_q \to \rho} & 0.221 & 0.023 & & \phantom{-}0.009 & -         &
\phantom{-}0.212 & 40.82 & 0.1 & (\ref{eq:dmex})\\ 
T_1^{B_q \to \rho} & 0.267 & 0.023 & & \phantom{-}0.897 & m^2_{1^-} & -0.629 & 38.04 & 0.1 &(\ref{eq:dp})\\ 
T_2^{B_q \to \rho} & 0.267 & 0.023 & & -     & -         & \phantom{-}0.267 & 38.59 & 2.3 &(\ref{eq:vmd})\\ 
\tilde{T}_3^{B_q \to \rho} & 0.267 & 0.023 & & \phantom{-}0.022 & -         & \phantom{-}0.246 & 40.88 & 0.1 & (\ref{eq:dmex})\\\rline
V^{B_s \to \rar{K}^*} & 0.311 & 0.026 & -0.43\relta_{a_1}   & \phantom{-}2.351 & m^2_{1^-} & -2.039 & 33.10 & 0.1 &(\ref{eq:dp})\\ 
A_0^{B_s \to \rar{K}^*} & 0.360 & 0.034 & -0.37\relta_{a_1} & \phantom{-}2.813 & m^2_{0^-} & -2.509 & 31.58 & 0.1 &(\ref{eq:dp})\\ 
A_1^{B_s \to \rar{K}^*} & 0.233 & 0.022 & -0.32\relta_{a_1} & - & - & \phantom{-}0.231 & 32.94 & 0.8 &(\ref{eq:vmd})\\ 
A_2^{B_s \to \rar{K}^*} & 0.181 & 0.025 & -0.30\relta_{a_1} & -0.011 & - & \phantom{-}0.192 & 40.14 & 0.1 &(\ref{eq:dmex})\\ 
T_1^{B_s \to \rar{K}^*} & 0.260 & 0.024 & -0.33\relta_{a_1} & \phantom{-}2.047 & m^2_{1^-} & -1.787 & 32.83 & 0.1 &(\ref{eq:dp})\\ 
T_2^{B_s \to \rar{K}^*} & 0.260 & 0.024 & -0.33\relta_{a_1} & - & - & \phantom{-}0.260 & 33.01 & 1.9 &(\ref{eq:vmd})\\
\tilde{T}_3^{B_s \to \rar{K}^*} & 0.260 & 0.024 
& -0.33\relta_{a_1} & \phantom{-}0.043 & - & \phantom{-}0.217 & 39.38 & 0.1 &(\ref{eq:dmex})\\  \rline
V^{B_q \to K^*} & 0.411 & 0.033 & \phantom{-}0.44\relta_{a_1}   & \phantom{-}0.923 & m^2_{1^-} & -0.511 & 49.40 & 0.0 &(\ref{eq:dp})\\ 
A_0^{B_q \to K^*} & 0.374 & 0.033 & \phantom{-}0.39\relta_{a_1} & \phantom{-}1.364 & m^2_{0^-} & -0.990 & 36.78 & 0.1 &(\ref{eq:dp})\\ 
A_1^{B_q \to K^*} & 0.292 & 0.028 & \phantom{-}0.33\relta_{a_1} & - & - & \phantom{-}0.290 & 40.38 & 1.0 &(\ref{eq:vmd})\\ 
A_2^{B_q \to K^*} & 0.259 & 0.027 & \phantom{-}0.31\relta_{a_1} & -0.084 & - & \phantom{-}0.342 & 52.00 & 0.2 &(\ref{eq:dmex})\\ 
T_1^{B_q \to K^*} & 0.333 & 0.028 & \phantom{-}0.34\relta_{a_1} & \phantom{-}0.823 & m^2_{1^-} & -0.491 & 46.31 & 0.0 &(\ref{eq:dp})\\ 
T_2^{B_q \to K^*} & 0.333 & 0.028 & \phantom{-}0.34\relta_{a_1} & - & - & \phantom{-}0.333 & 41.41 & 2.5 &(\ref{eq:vmd})\\ 
\tilde{T}_3^{B_q \to K^*} & 0.333 & 0.028 &
\phantom{-}0.34\relta_{a_1} & -0.036 & - &\phantom{-} 0.368 & 48.10 &
0.1 & (\ref{eq:dmex})\\  \rline
V^{B_q \to \rmega} & 0.293 & 0.029 &   & \phantom{-}1.006 & m^2_{1^-}& -0.713 & 37.45 & 0.1 &(\ref{eq:dp})\\ 
A_0^{B_q \to \rmega} & 0.281 & 0.030 & & \phantom{-}1.321 & m^2_{0^-} & -1.040 & 34.47 & 0.1 &(\ref{eq:dp})\\ 
A_1^{B_q \to \rmega} & 0.219 & 0.024 & & - & - & -0.217 & 37.01 & 1.1 & (\ref{eq:vmd})\\ 
A_2^{B_q \to \rmega} & 0.198 & 0.023 & & \phantom{-}0.006 & - & \phantom{-}0.192 & 41.24 & 0.1 &(\ref{eq:dmex})\\ 
T_1^{B_q \to \rmega} & 0.242 & 0.021 & & \phantom{-}0.865 & m^2_{1^-} & -0.622 & 37.19 & 0.1 &(\ref{eq:dp})\\ 
T_2^{B_q \to \rmega} & 0.242 & 0.021 & & - & - & \phantom{-}0.242 & 37.95 & 2.1 &(\ref{eq:vmd})\\ 
\tilde{T}_3^{B_q \to \rmega} & 0.242 & 0.021 & & \phantom{-}0.023 & - & \phantom{-}0.220 & 40.87 & 0.1& (\ref{eq:dmex})\\  \rline
V^{B_s \to \rhi} & 0.434 & 0.035 &     & \phantom{-}1.484 & m^2_{1^-} & -1.049 & 39.52 & 0.1& (\ref{eq:dp})\\ 
A_0^{B_s \to \rhi} & 0.474 & 0.033 &   & \phantom{-}3.310 & m^2_{0^-} & -2.835 & 31.57 & 0.1& (\ref{eq:dp})\\ 
A_1^{B_s \to \rhi} & 0.311 & 0.030 &   & - & - & \phantom{-}0.308 & 36.54 & 1.0& (\ref{eq:vmd})\\ 
A_2^{B_s \to \rhi} & 0.234 & 0.028 &   & -0.054 & - & \phantom{-}0.288 & 48.94 & 0.2& (\ref{eq:dmex})\\ 
T_1^{B_s \to \rhi} & 0.349 & 0.033 &   & \phantom{-}1.303 & m^2_{1^-} & -0.954 & 38.28 & 0.1& (\ref{eq:dp})\\ 
T_2^{B_s \to \rhi} & 0.349 & 0.033 &   & - & - & \phantom{-}0.349 & 37.21 & 2.4& (\ref{eq:vmd})\\ 
\tilde{T}_3^{B_s \to \rhi} & 0.349 & 0.033 &   & \phantom{-}0.027 & - & \phantom{-}0.321 & 45.56 & 0.1& (\ref{eq:dmex})\\  \rline
\end{array}
$$
\vskip-1pt
\addtolength{\arraycolsep}{-3pt}
\caption[]{Fits for the form factors 
valid for general $q^2$. Columns 2 to 4 give
  the results of Tab.~\ref{tab:basic} for $q^2=0$, including the
  errors $\Delta_{\rm tot}$ and $\Delta_{a_1}$.
The remaining colums give the fit parameters. 
Note that we  fit
the form factor $\tilde{T}_3$, defined in Eq.~(\ref{eq:eq}), instead of
  $T_3$. The fit formulas to use are given in the last column, the
  masses $m_R$ are given in Tab.~\ref{tab:poles}.
The penultimate column gives the fit error
$\delta$ as defined in Eq.~\eqref{eq:delta}.}\label{tab:fits}
\end{table} 

The dominant poles at $q^2=m_R^2$ correspond to resonances with
quantum numbers $J^P=1^-$ for $V$ and $T_1$, $0^-$ for $A_0$  and
$1^+$ for $A_{1,2,3}$ and $T_{2,3}$, $\widetilde{T}_3$. As discussed
in Sec.~\ref{sec:2}, not all these form factors are independent, and
the question arises which ones to fit to the above equation -- or any
similar formula -- and which ones to define in terms of the others. As
Eq.~(\ref{eq:dp}) contains two explicit poles, we decide the above
question in favor of the form factors with the steepest increase in
$q^2$, which means that the independent form factors are $V$,
$A_{0,1,2}$ and $T_{1,2}$, $\tilde{T}_3$, whereas $T_3$ and $A_3$ are
the dependent ones, defined as in Eqs.~(\ref{eq:A30}) and (\ref{eq:T3tilde}).

The values of the
resonance masses $m_R$ in (\ref{eq:dp}) 
are known from experiment for $0^-$ and $1^-$
in the $B_q$ channel and $0^-$ in the $B_s$ channel;
the other masses are obtained using heavy quark symmetry relations
\cite{Bardeen},
the numerical values are collected in Tab.~\ref{tab:poles}.

We shall use fits to Eq.~(\ref{eq:dp}) for the form factors $V$, $A_0$
and $T_1$, where the lowest pole $m_R^2$ lies well below the
multiparticle threshold $(m_{B_{q,s}} + m_{\pi,K})^2$.
\begin{table}[tbp]
\renewcommand{\arraystretch}{1.2}\addtolength{\arraycolsep}{3pt}
$$
\begin{array}{|l|cccc|}
\hline
      & 0^- & 0^+ & 1^- & 1^+ \\ \hline
B_q & 5.28  & 5.63  & 5.32  & 5.68  \\
B_s & 5.37  & 5.72  & 5.42  & 5.77  \\  
\hline
\end{array}
$$
\caption[]{$B$ meson masses in units GeV, taken from 
Ref.~\cite{Bardeen}.}\label{tab:poles}
\end{table}
If, on the other hand, 
the lowest physical pole lies sufficiently close to the
multiparticle threshold $t_0$ or even above it, then it may be impossible to
``resolve'' the poles from a low-$q^2$ ``perspective''. In this case
it is more appropriate to expand the form factor to second order around the
pole, yielding
\begin{equation}\label{eq:dmex}
F(q^2)=\frac{r_1}{1-q^2/m_{\rm fit}^2}+\frac{r_2}{(1-q^2/m_{\rm fit}^2)^2}\,,
\end{equation}
with the three parameters $r_{1,2}$ and $m_{\rm fit}$. This is the fit
formula we shall use for the axialvector form factors, in particular 
$A_2$ and $\tilde{T}_3$. For $A_1$ and $T_2$,
on the other hand, the residue of the double pole in $m_{\rm fit}$
turns out to be
extremely small, so that it can be dropped and one is back to the VMD formula
\begin{equation}\label{eq:vmd}
F(q^2)= \frac{r_2}{1-q^2/m_{\rm fit}^2}\,,
\end{equation}
albeit with an effective pole mass $m_{\rm fit}$ unrelated to any resonance.

The fits of the LCSR results to the above formulas are collected in
Tab.~\ref{tab:fits}; they  
differ from the LCSR results obtained for $q^2\leq 14\,{\rm
  GeV}^2$, by no more than 2.5\%. In Tab.~\ref{tab:fits} we
indicate the ``quality'' of the fit by  $\delta$,
which is the maximum deviation of the fit relative to the mean value
of the form factor in percent and defined as
\begin{equation}\label{eq:delta}
\delta = 100\:\frac{\sum_t|f(t)-f^{\rm fit}(t)|}{\sum_{t}|f(t)|},
\end{equation}
where the sum runs over $t\in\{ 0,0.5,1,\dots,14\}$. 

\begin{figure}[tb]
$$\epsfxsize=0.45\textwidth\epsffile{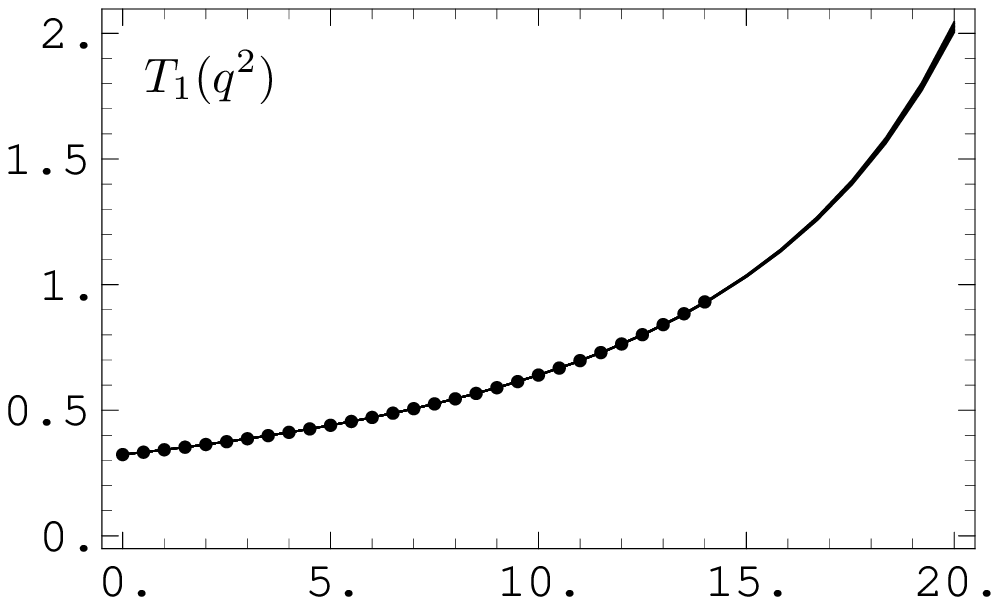} \quad 
\epsfxsize=0.45\textwidth\epsffile{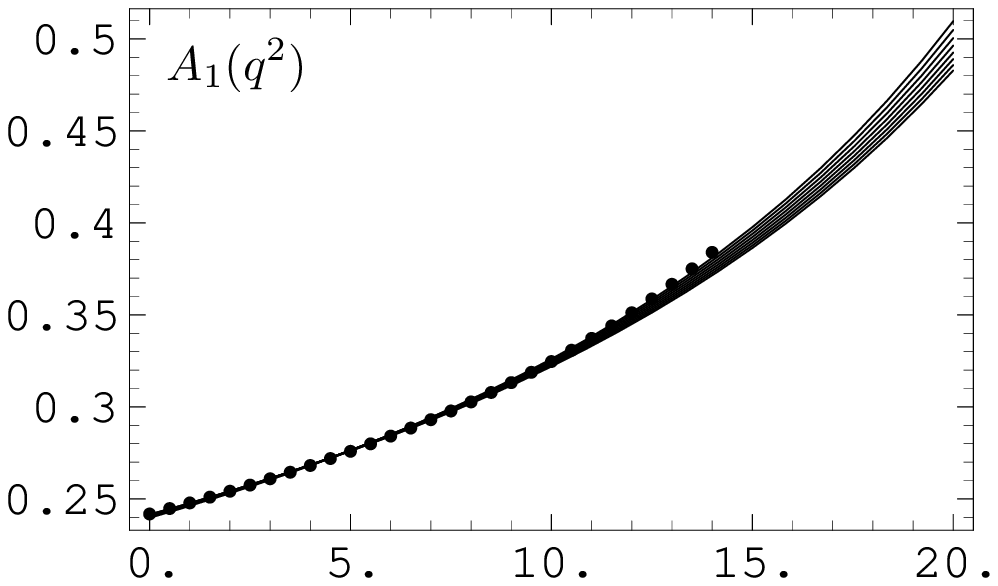}$$
\caption[]{Comparison of the consistency of fits of
  $T_{1}^{B_q\to\rho}$ and $A_{1}^{B_q\to\rho}$obtained for
  different values of $q^2_{\rm LCSR, max}$. Dots: LCSR results for
  $q^2\leq 14\,{\rm GeV}^2$. Lines: fits according 
to Eqs.~(\ref{eq:dp}) for $T_1$ and (\ref{eq:vmd}) for $A_1$, for
  $q^2_{\rm LCSR, max}$ between 7 and $14\,{\rm GeV}^2$. The maximum
  discrepancy between the fit-results at $q^2=20\,{\rm GeV}^2$ is
  2\% for $T_1$ and 5\% for $A_1$.}\label{fig:fitqu}
\end{figure}

We have also tried fits to 
the two pole ansatz (\ref{eq:dp}) without fixing one of the
masses. In this case the lowest pole is fitted to lie below the actual
resonance pole, by up to  
$1.5\,{\rm GeV}^2$. Given the fact that LCSRs are
valid for small $q^2$ far away from the pole, one cannot expect them 
to resolve its position with perfect accuracy. Nonetheless we take it
as an indication for the consistency of our approach that the double
pole formula with unrestricted pole positions gives results
that agree qualitatively with those from the restricted fits.
We also have checked the dependence of the fits on the maximum value
of $q^2_{\rm max, LCSR}$ up to which LCSR results are included into
the fit. It turns out that the fits are very robust against lowering
$q^2_{\rm max, LCSR}$; lowering it from 14$\,{\rm GeV}^2$ 
to 7$\,{\rm GeV}^2$  changes the fitted values at 20$\,{\rm GeV}^2$ by at
most 8\%, 
$T_2$ being the odd one sticking out. In Fig.~\ref{fig:fitqu} we show the
effects of a change of $q^2_{\rm max, LCSR}$ on $T_1^{B\to\rho}$ and
$A_1^{B\to\rho}$.

Let us now turn to a  consistency check of our fits. One can express
the residues of
$V$, $T_1$ and $A_0$ for $B\to\rho$ in terms of decay constants and
strong couplings as follows:
\begin{equation}\label{eq:residues}
r_1^V = \frac{m_B+m_V}{2 m_B} f_{B^*}\, g_{BB^*\rho},\quad
r_1^{T_1} = \frac{f_{B^*}^T}{2}\,  g_{BB^*\rho},\quad
r_1^{A_0} = \frac{f_B}{2 m_V}\, g_{BB\rho},
\end{equation}
where $f_{B^*}^T$ is the tensor coupling of the $B^*$ meson
defined in the same way as  light vector tensor couplings,
Eq.~\eqref{eq:frp}. 
$f_B$ has been discussed in Sec.~\ref{sec:4.2}; its value is about 
$200\, {\rm MeV}$ and we expect $f_{B^*}$ and $f^T_{B^*}$ to be of about 
the same size.
The values of the strong couplings $g_{BB\rho}$ and $g_{BB^*\rho}$ are
more controversial as discussed below.
As a first check, consider the $g$-independent ratio
\begin{equation}\label{eq:expecto_patronum}
\alpha \equiv \frac{r_1^V}{r_1^{T_1}} = \frac{m_B+m_V}{m_{B^*}}\,
\frac{f_{B^*}}{f_{B^*}^T} \sim 1.14.
\end{equation}
The fitted values of $r_1$ are collected in Tab.~\ref{tab:Vergleich}
and yield $\alpha_{\rm fit} = 1.16$ --- 
very close to (\ref{eq:expecto_patronum}). For $r_1$ fitted using
parameter sets 1 and 3 we find $\alpha=1.16$ and 1.17, respectively.
Any further check of the fitted $r_1$ requires information on the
couplings $g_{BB^{(*)}\rho}$, which have been calculated from both
LCSRs \cite{China} and within the
constituent quark meson (CQM) model \cite{Italia} --- with
significantly different results. The situation resembles that for 
$g_{DD^*\pi}$, where LCSR determinations are typically by a factor 2
smaller than lattice and CQM calculations 
\cite{gDDpi_lat}. For this coupling there actually  exists  an experimental
measurement by CLEO \cite{CLEO2}, which agrees with the lattice and
CQM determinations, but disagrees with LCSRs. For the corresponding $B$
coupling $g_{BB^*\pi}$ there is no experimental measurement, as the
decay $B^*\to B\pi$ is forbidden by phase space, but one can  use
heavy quark scaling to obtain $g_{BB^*\pi}$ from the measured
$g_{DD^*\pi}$ and compare it with the corresponding theoretical
predictions. It turns out that again lattice and CQM calculations are
favored, whereas the LCSR calculation gives a too small result, which
can be understood following the discussion in Ref.~\cite{possible}.
The recent LCSR  determination \cite{China} 
has up-to-date input parameters and they get
from a tree-level analysis
\begin{equation}
\label{eq:China}
g_{BB\rho} = 5.37,\quad g_{BB^*\rho} = 5.70 \,{\rm GeV}^{-1}\,.
\end{equation} 
For pseudoscalar mesons, NLO calculations have consistently yielded
smaller values than tree level determinations, 
cf.\ Ref.~\cite{gBBpi}, which, 
if true also for the $\rho$, would widen  
the gap between the results from different methods even further.
The CQM-model predictions are \cite{Italia}:
\begin{eqnarray}
\label{eq:Italia}
g_{BB\rho}   &=& -\sqrt{2} \beta \frac{m_\rho}{f_\pi} = 
7.2 \quad {\rm with } \quad \beta = -0.86, \\ 
g_{BB^*\rho} &=& \sqrt{8} \lambda \,
\sqrt{\frac{m_{B^*}}{m_B}} \frac{m_\rho}{f_\pi} = 10.0 \: {\rm GeV}^{-1}
\quad {\rm with } \quad \lambda = 0.6\, {\rm GeV}^{-1}.\label{eq:Italia2}
\end{eqnarray}
It is hard for us to judge on the validity of this approach, 
but as far as we understand the model is further based on empirical success.
In Tab.~\ref{tab:Vergleich} we compare the
residues for the $B \to \rho$ transition
as obtained from our fits, Tab.~\ref{tab:fB}, to their values
given in Eq.~\eqref{eq:residues}, using the couplings \eqref{eq:China},
\eqref{eq:Italia} and \eqref{eq:Italia2}. 
\begin{table}
\renewcommand{\arraystretch}{1.2}
\addtolength{\arraycolsep}{3pt}
$$
\begin{array}{|l|ccc|}\hline
& \mbox{our fit} & \mbox{LCSR \cite{China}} & 
\mbox{CQM \cite{Italia}} \\ \hline
r_1^V      & 1.05  & 0.65 & 1.14 \\
r_1^{T_1}  & 0.90  & 0.57 & 1.00 \\
r_1^{A_0}  & 1.53  & 0.70 & 0.94 \\
\hline
\end{array}
\renewcommand{\arraystretch}{1.0}
\addtolength{\arraycolsep}{-3pt}
$$
\caption[]{Residues of the lowest-lying pole for  
$V^{B_q\to\rho}$, $T_1^{B_q\to\rho}$ and $A_0^{B_q\to\rho}$ obtained
  from our fits
as compared to Eq.~\eqref{eq:residues} with input values from LCSR 
and CQM determinations.}\label{tab:Vergleich} 
\end{table}
For $V$ and $T_1$ with  a $1^-$ pole 
the CQM residues are about $10\%$ larger, 
and the LCSR about $40\%$
lower than the fitted values. As discussed above, the LCSR results are
expected to fall short of the real values, so this is an excellent
confirmation of our results.
The $A_0$ form factor shows some discrepancy which may indicate that 
either the estimate of the $g_{BB\rho}$ coupling is too low or that 
the second pole in the fit, $m_{\rm fit}^2 \approx 33\,{\rm GeV}^2$, is 
too close to the resonance pole to allow a clean determination of
its residue. Taken altogether, however, the agreement of our fitted
results to that of independent calculations is an excellent confirmation or
our results.

\section{Summary \& Conclusions}

In this paper we present a thorough and careful examination of the
predictions of QCD sum rules on the light-cone for the form factors
of $B_q$ and $B_s$ transitions to $\rho$, $\omega$, $K^*$ and $\phi$.
Our main results for zero momentum transfer 
$q^2=0$ are collected in Tab.~\ref{tab:basic},
those for general $q^2$ in Tab.~\ref{tab:fits}.

The present analysis is a sequel of our work on
$B\to\,$pseu\-do\-sca\-lar form factors, Ref.~\cite{BZ}, and an
extension of the previous work of one of us on $B\to\,$vector form
factors, Ref.~\cite{BB98}. It improves upon the latter by
\begin{itemize}
\item including predictions for all form factors of $B_{q,s}$ transitions
  to $O(\alpha_s)$ accuracy for twist-2 and 3 2-particle
  contributions;
\item a more sophisticated method for fixing sum rule specific
  parameters, cf.\ Sec.~\ref{sec:4.2};
\item implementing recently developed new models for the dominant
  nonperturbative vector meson contributions, the twist-2 vector meson
  distribution amplitudes, cf.\ Ref.~\cite{angi};
\item the possibility to implement future updates of some hadronic 
parameters in a straightforward way, cf.\ App.~\ref{app:0};
\item a careful assessment of uncertainties at zero momentum transfer,
  cf.\ Sec.~\ref{sec:4.4};
\item a parametrization of the $q^2$-dependence of form factors valid
  in the full physical regime of momentum transfer, cf.\
  Sec.~\ref{sec:4.5};
\item a variety of consistency checks for the robustness of the $q^2$-fits
  and their numerical results.
\end{itemize}

The accuracy of our results is limited, on the one hand, by the
uncertainty of hadronic input parameters and, on the other hand, by
the systematic uncertainty induced by the fact that QCD sum rules on
the light-cone are an approximative method. The uncertainty due to
the variation of only the sum rule specific parameters is about 
7\%, which cannot be reduced any further and hence sets the minimum theoretical
uncertainty that can be achieved within this method. An equally large
theoretical uncertainty is induced by hadronic parameters and can, in
principle, be improved upon. We quote in particular the tensor
couplings $f_V^T$ of vector mesons, which presently come with the
rather large error quoted in Tab.~\ref{tab:fV}. Improvement should be
possible by  dedicated lattice calculations, a first example of
which is Ref.~\cite{lattfT}. Another relevant hadronic parameter is
$\Delta$, the first inverse moment of the twist-2 vector meson
distribution amplitudes, as defined in Sec.~\ref{sec:4.3}. 
We have inferred a likely range for this
parameter for $\rho$ and $\omega$ mesons from the known experimental
constraints on $\Delta(\pi)$, and further determined a range for
$\Delta(K^*)$ and $\Delta(\phi)$ from the observed decrease, within QCD
sum rule calculations, of the
second Gegenbauer moment $a_2$ with increasing meson mass. Comparing the 
theoretical errors collected in Tab.~\ref{tab:basic} with 
the global theoretical uncertainty $\sim 15\%$ quoted in our previous 
publication \cite{BB98}, we have achieved a reduction to about 10\%. This
is partially due to a reduction of the uncertainties of the hadronic input
parameters, in particular $m_b$, and partially due to a refinement
of the assessment of sum rule specific uncertainties as discussed in
Sec.~\ref{sec:4.2}. Any future reduction of the total uncertainty will
depend on more accurate
determinations of $\Delta$, which are absolutely essential not only for
light-cone sum rule calculations, but also for exploiting the full
potential of QCD factorisation formulas for nonleptonic exclusive B
decays \cite{QCDfac}. We take this occasion to urge lattice
practitioners to take up the challenge and develop new and ingenious
methods to tackle this problem --- or just give us an accurate 
value of $a_2$, which would already be a big step forward.

The prospects for future direct determinations of $B\to V$ form factors from
lattice calculations do appear a bit clouded. On the one hand, there
are two recent studies, by the SPQcdR and UKQCD collaborations,
Ref.~\cite{early}, using an improved  
Wilson action and the quenched approximation. The $b$ quarks are fully
relativistic and have typical masses of about 2 to 3~GeV, so they need
to be extrapolated to the physical $b$ quark mass. On the other hand,
we conclude from \cite{davies} that an unquenched calculation in NRQCD
is not really on the menu, which, as far as
we understand, is due to an improvement in the treatment of light
quark masses on the lattice, causing the $\rho$ and other vector
mesons to become instable particles without a pronounced plateau in the
fall-off of the correlation function, and essentially prevents  a precise
determination of their properties from lattice. We do not pretend to
be sufficient experts in LQCD to be able to meaningfully comment on
these issues, but remain hopeful that the situation will be clarified
in due course.

We have calculated all form factors for $0\leq q^2\leq 14\,{\rm
  GeV}^2$; the upper bound on $q^2$ is due to the limitations of the
  light-cone expansion which requires the final-state meson to have
  energies $E\gg \Lambda_{\rm QCD}$: for $q^2_{\rm max} = 14\,{\rm
  GeV}^2$ the meson energy is $E = 1.3\,$GeV. In order to facilitate the
  use of our results we have given, in Sec.~\ref{sec:4.5},
  Eqs.~(\ref{eq:dp}), (\ref{eq:dmex}) and (\ref{eq:vmd}), simple 
  parametrisations that include the main features of the analytical
  properties of the form factors and are valid in the full physical
  regime $0\leq q^2\leq (m_B-m_V)^2$. The corresponding results for
  our preferred set of input parameters are given in Tab.~\ref{tab:fits}. 
We have checked that the fit results are fairly insensitive to the maximum
  value of $q^2$ included, and that reducing the latter to e.g.\ $7\,{\rm
  GeV}^2$ changes the extrapolated values of the form factors at
  $q^2=20\,{\rm GeV}^2$ by typically only 1 to 2\%, and by 8\% at most
  (for $T_2$).

In Sec.~\ref{sec:intro} we  mentioned factorisation formulas for form
factors derived in SCET, Ref.~\cite{SCET,fuck,hill}, which in particular
imply certain (heavy quark) symmetry relations. Since the
objective of this paper was to provide numerical results, ready for
use in phenomenological applications, we did not discuss the question
whether and to what extent our results fulfill these relations, nor
the size of symmetry-breaking corrections. A previous study of the
corresponding effect in $B\to\,$pseudoscalar decays has indicated that
such corrections are likely to be nonnegligible \cite{yetanotherone}. 
We plan to come
back to these points in a future publication.

\section*{Acknowledgements}

R.Z.is supported by the Swiss National Science Foundation and
would like to thank Antonio Polosa for correspondence on Ref.~\cite{Italia}.

\appendix

\section*{Appendix}
\setcounter{equation}{0}
\renewcommand{\theequation}{A.\arabic{equation}}
\renewcommand{\thetable}{\Alph{table}}
\setcounter{section}{0}
\setcounter{table}{0}

\begin{table}[tbp]
\renewcommand{\arraystretch}{1.3}
\begin{center}
\begin{tabular}{|c|cc||c|cc|}
\hline
$F(0)$ & $F^L$ & $F^T$ & $F(0)$ & $F^L$ & $F^T$ \\
\hline
$V^{B_q \to \rho}$ & 0.1092 & 0.2139 & $A_0^{B_q \to \rho}$ & 0.2036 & 0.0990\\ 
$A_1^{B_q \to \rho}$ & 0.0867 & 0.1552 & $A_2^{B_q \to \rho}$ & 0.0467 & 0.1743\\ 
$T_1^{B_q \to \rho}$ & 0.1034 & 0.1641 & $T_3^{B_q \to \rho}$ & 0.0303 & 0.1455 \\ 
\hline
$V^{B_s \to \rar{K}^*}$ & 0.1275 & 0.2289 & $A_0^{B_s \to \rar{K}^*}$ & 0.2469 & 0.1532\\ 
$A_1^{B_s \to \rar{K}^*}$ & 0.1022 & 0.1641 & $A_2^{B_s \to \rar{K}^*}$ & 0.0445 & 0.1684\\ 
$T_1^{B_s \to \rar{K}^*}$ & 0.1215 & 0.1737 & $T_3^{B_s \to \rar{K}^*}$ & 0.0211 & 0.1339\\ 
\hline
$V^{B_q \to K^*}$ & 0.1415 & 0.2234 & $A_0^{B_q \to K^*}$ & 0.2071 & 0.1269\\ 
$A_1^{B_q \to K^*}$ & 0.1034 & 0.1545 & $A_2^{B_q \to K^*}$ & 0.0614 & 0.1658\\ 
$T_1^{B_q \to K^*}$ & 0.1301 & 0.1665 & $T_3^{B_q \to K^*}$ & 0.0436 & 0.1386\\
\hline 
$V^{B_q \to \rmega}$ & 0.1048 & 0.1884  & $A_0^{B_q \to \rmega}$ & 0.1967 & 0.0838\\ 
$A_1^{B_q \to \rmega}$ & 0.0834 & 0.1357 & $A_2^{B_q \to \rmega}$ & 0.0443 & 0.1536\\ 
$T_1^{B_q \to \rmega}$ & 0.0984 & 0.1440 & $T_3^{B_q \to \rmega}$ & 0.0288 & 0.1264\\ 
\hline
$V^{B_s \to \rhi}$ & 0.1594 & 0.2748 & $A_0^{B_s \to \rhi}$ & 0.2647 & 0.2098\\ 
$A_1^{B_s \to \rhi}$ & 0.1226 & 0.1884 & $A_2^{B_s \to \rhi}$ & 0.0558 & 0.1784\\ 
$T_1^{B_s \to \rhi}$ & 0.1469 & 0.2019 & $T_3^{B_s \to \rhi}$ & 0.0316 & 0.1433\\ 
\hline
\end{tabular}
\end{center}
\caption[]{Contributions  in $f^L$ and
  $f^T$ to the form factors at $q^2=0$. The numbers correspond to
  the central values of parameter set 2, i.e.\
  $m_b=4.8\,$GeV. $T_2(0)$ follows from  $T_1(0)=T_2(0)$.}\label{tab:fLfT1}
\renewcommand{\arraystretch}{1.3}\addtolength{\arraycolsep}{3pt}
$$
\begin{array}{|c|cc||c|cc|}
\hline
F(0) & F^{L,a_1} & F^{T,a_1} & F(0) & F^{L,a_1} & F^{T,a_1}\\\hline
V^{B_s \to \rar{K}^*} & -0.0057 & -0.0396 & A_0^{B_s \to\rar{K}^*} &
-0.0398 & \phantom{-}0.0024\\ 
A_1^{B_s \to \rar{K}^*} & -0.0057 & -0.0276 & A_2^{B_s \to
  \rar{K}^*} & \phantom{-}0.0079 & -0.0394 \\ 
T_1^{B_s \to \rar{K}^*} & -0.0056 & -0.0297 & T_3^{B_s \to
  \rar{K}^*} & \phantom{-}0.0104 & -0.0293 \\ \hline
V^{B_q \to K^*} & \phantom{-}0.0060 & \phantom{-}0.0403 & A_0^{B_q
  \to K^*} & \phantom{-}0.0403 & -0.0001 \\ 
A_1^{B_q \to K^*} & \phantom{-}0.0059 & \phantom{-}0.0281 & A_2^{B_q
  \to K^*} & -0.0080 & \phantom{-}0.0395 \\ 
T_1^{B_q \to K^*} & \phantom{-}0.0059 & \phantom{-}0.0303 &
T_3^{B_q \to K^*} & -0.0103 & \phantom{-}0.0299 \\\hline 
\end{array}
$$
\renewcommand{\arraystretch}{1}
\caption[]{Contributions of $a_1$ to the form factors at $q^2=0$. 
Parameters like in  Tab.~\ref{tab:fLfT1}.}\label{tab:fLfT2}
\end{table} 

\section{\boldmath Form Factors For Different $f_V^{(T)}$ and 
$a_1(K^*)$}\label{app:0}

The form factors can be written as sum of two contributions which are
proportional to the vector meson's vector coupling $f_V\equiv f^L$ and
the tensor coupling $f_V^T\equiv f^T$, respectively. The uncertainties
of these parameters as tabled in Tab.~\ref{tab:fV} are nonnegligible,
but amenable to future improvement by e.g.\ lattice calculations, cf.\ 
Ref.~\cite{lattfT}. The same applies to $a_1(K^*)$ which also comes with a
considerable uncertainty, cf.\ Eq.~(\ref{a1}). 
In order to allow the adjustment of our results to improved
determinations of these parameters, we write the generic form factor
$F$ as
\begin{equation}\label{eq:fsplit}
F = \hat{f}^L(F^L+\hat{a}^L_1\,
F^{L,a_1})+\hat{f}^T(F^T+ \hat{a}^T_1\, F^{T,a_1}) \quad ,
\end{equation}
where the hatted quantities are normalized to the central values used
in our calculation, i.e.\ the couplings of Tab.~\ref{tab:fV} and
$a_1^T\equiv a_1^\perp$, $a_1^L\equiv a_1^\parallel$ as given in
(\ref{a1}). For instance $\hat{f}^L_\rho = f_\rho/(205\,{\rm MeV})$. 
Note that $\hat{a}_1(K^*)\equiv \hat{a}_1(\bar K^*)$ and
that hatted quantities are trivially invariant under LO scaling.
$F^L$ and $F^T$ are collected, for $q^2=0$, in Tab.~\ref{tab:fLfT1}
and $F^{a_1}$ in Tab.~\ref{tab:fLfT2}. To give an example, for $A_0^{B_s \to
  \bar{K}^*}$ we obtain
\begin{eqnarray*}
A_0^{B_s \to \bar{K}^*}(0) &=& \frac{f_{K^*}}{217\,\text{\small MeV}} \left(0.2469 -          
\frac{a_1^{K^*}(\text{\small 1 \rm GeV})}{0.1}\, 0.0398 \right)\\
&&{}+ \frac{f^T_{K^*}(\text{\small 1 \rm GeV}) }
{170\,\text{\small MeV}}\left(0.1532 +          
\frac{a_1^{K^*}(\text{\small 1 \rm GeV})}{0.1}\, 0.0024\right),
\end{eqnarray*}
which, choosing the central values of the couplings and $a_1$, yields
0.3627, in agreement with Tab.~\ref{tab:basic}.

\section{Distribution Amplitudes}\label{app:A}

In this appendix we collect explicit expressions for some of the 
twist-3 and 4 DAs that
enter the LCSRs. These expressions are well-known and have been
taken from Ref.~\cite{wavefunctions}. The twist-2 DAs have already
been discussed in Sec.~\ref{sec:4.3}. We also motivate and justify the
use of models for DAs based on a truncated conformal expansion.

Before defining the DAs, we introduce the light-like vectors
in which they are expressed. We denote the meson momentum by $P_\mu$
(with $P^2=m_V^2$) and the separation between fields in a nonlocal
operator by $x_\mu$ (with $x^2$ close to 0)
and introduce light-like vectors $p$ and $z$ such that 
\begin{equation}
p_\mu = P_\mu-\frac{1}{2}z_\mu \frac{m^2_V}{pz}\,, \qquad
z_\mu  =  x_\mu - P_\mu \left[ xP - \sqrt{(xP)^2 - x^2 m_V^2}\right]/m_V^2.
\end{equation}  
The meson polarization vector $e^{(\lambda)}_\mu$ is decomposed into 
projections onto the two light-like vectors and the orthogonal plane 
as
\begin{equation}
 e^{(\lambda)}_\mu = \frac{(e^{(\lambda)} z)}{pz}
\left( p_\mu -\frac{m^2_V}{2pz} z_\mu \right)+e^{(\lambda)}_{\perp\mu}. 
\label{polv}
\end{equation} 
We also need the projector onto the directions orthogonal to $p$ and $z$:
\begin{equation}
       g^\perp_{\mu\nu} = g_{\mu\nu} -\frac{1}{pz}(p_\mu z_\nu+ p_\nu z_\mu).
\end{equation}
The dual gluon field strength
tensor is defined as $\widetilde{G}_{\mu\nu} =
\frac{1}{2}\epsilon_{\mu\nu \rho\sigma} G^{\rho\sigma}$. 
We use the standard Bjorken-Drell
convention \cite{BD65} for the metric tensor 
and the Dirac matrices; in particular
$\gamma_{5} = i \gamma^{0} \gamma^{1} \gamma^{2} \gamma^{3}$,
and the Levi-Civita tensor $\epsilon_{\mu \nu \lambda \sigma}$
is defined as the totally antisymmetric tensor with $\epsilon_{0123} = 1$.
This convention differs in sign by the one of Itzykson/Zuber \cite{IZ}
used in some programming packages, e.g.\ {\sc FeynCalc}.
We use a sign-convention for the strong coupling $g$ where the 
covariant derivative is defined as
$D_{\mu}= \partial_{\mu} - igA_{\mu}$ and hence the Feynman-rule for
$qqg$ vertices is $+ig\gamma_\mu$.

Let us also clarify the treatment of SU(3)-breaking effects in
DAs. SU(3) breaking occurs in three different ways:
\begin{itemize}
\item the contribution of odd Gegenbauer-moments $a_{1,3,\dots}$ to
  the DAs of the $K^*$;
\item a difference in the values of the couplings $f_V^{(T)}$,
   the even Gegenbauer-moments
  $a_2^{\rho}\neq a_2^{K^*}$ and 3-particle matrix elements;
\item the modification of relations between DAs by terms in
  $m_{q_1}\pm m_{q_2}$.
\end{itemize}
We will take into account the first effect wherever it occurs, except
for terms in $O(\delta^2)$, the reason being that the structure of
$\delta^2$ terms is very involved and there are yet unknown
contributions in $m_V^2 a_1^\perp$ induced by 3-particle
twist-4 DAs. The second effect is taken into account for the decay
constants and parametrized by the dependence of
the form factors on the parameters $\Delta$, as discussed in
Sec.~\ref{sec:4.3}; we do not include SU(3) breaking for the
3-particle matrix elements as information on these effects is
virtually nonexistant. The third effect is taken into account at
$O(\delta \alpha_s^0,\delta \alpha_s)$, i.e.\ for the chiral-even DAs
$g_\perp^{(a,v)}$. It does not occur at $O(\delta^0)$ and the
corresponding terms are unknown at $O(\delta^2)$. 

The 2-particle DAs have been defined in Eqs.~(\ref{eq:OPEvector}) to
(\ref{eq:OPEx}). Up to twist-4 and $O(\delta^2)$, there are seven 
chiral-odd 3-particle DAs which can be defined as \cite{wavefunctions}
\begin{eqnarray}
\lefteqn{\langle 0|\bar q_2(z) \sigma_{\alpha\beta}
         gG_{\mu\nu}(vz)
         q_1(-z)|V(p,\lambda)\rangle \ =}\hspace*{1.6cm}\nonumber\\
&=& f_{V}^T m_{V}^2 \frac{e^{(\lambda)} z }{2 (p  z)}
    [ p_\alpha p_\mu g^\perp_{\beta\nu}
     -p_\beta p_\mu g^\perp_{\alpha\nu}
     -p_\alpha p_\nu g^\perp_{\beta\mu}
     +p_\beta p_\nu g^\perp_{\alpha\mu} ]
      {\cal T}(v,pz)
\nonumber\\
&&{}+ f_{V}^T m_{V}^2
    [ p_\alpha e^{(\lambda)}_{\perp\mu}g^\perp_{\beta\nu}
     -p_\beta e^{(\lambda)}_{\perp\mu}g^\perp_{\alpha\nu}
     -p_\alpha e^{(\lambda)}_{\perp\nu}g^\perp_{\beta\mu}
     +p_\beta e^{(\lambda)}_{\perp\nu}g^\perp_{\alpha\mu} ]
      T_1^{(4)}(v,pz)
\nonumber\\
&&{}+ f_{V}^T m_{V}^2
    [ p_\mu e^{(\lambda)}_{\perp\alpha}g^\perp_{\beta\nu}
     -p_\mu e^{(\lambda)}_{\perp\beta}g^\perp_{\alpha\nu}
     -p_\nu e^{(\lambda)}_{\perp\alpha}g^\perp_{\beta\mu}
     +p_\nu e^{(\lambda)}_{\perp\beta}g^\perp_{\alpha\mu} ]
      T_2^{(4)}(v,pz)
\nonumber\\
&&{}+ \frac{f_{V}^T m_{V}^2}{pz}
    [ p_\alpha p_\mu e^{(\lambda)}_{\perp\beta}z_\nu
     -p_\beta p_\mu e^{(\lambda)}_{\perp\alpha}z_\nu
     -p_\alpha p_\nu e^{(\lambda)}_{\perp\beta}z_\mu
     +p_\beta p_\nu e^{(\lambda)}_{\perp\alpha}z_\mu ]
      T_3^{(4)}(v,pz)
\nonumber\\
&&{}+ \frac{f_{V}^T m_{V}^2}{pz}
    [ p_\alpha p_\mu e^{(\lambda)}_{\perp\nu}z_\beta
     -p_\beta p_\mu e^{(\lambda)}_{\perp\nu}z_\alpha
     -p_\alpha p_\nu e^{(\lambda)}_{\perp\mu}z_\beta
     +p_\beta p_\nu e^{(\lambda)}_{\perp\mu}z_\alpha ]
      T_4^{(4)}(v,pz),\hspace*{1cm}
\label{eq:T3}\\
\lefteqn{\langle 0|\bar q_2(z)
         gG_{\mu\nu}(vz)
         q_1(-z)|V(p,\lambda)\rangle
\ =\ i f_{V}^T m_{V}^2
 [e^{(\lambda)}_{\perp\mu}p_\nu-e^{(\lambda)}_{\perp\nu}p_\mu] S(v,pz),}
\hspace*{1.6cm}\nonumber\\
\lefteqn{\langle 0|\bar q_2(z)
         ig\widetilde G_{\mu\nu}(vz)\gamma_5
         q_1(-z)|V(p,\lambda)\rangle
\ =\ i f_{V}^T m_{V}^2
 [e^{(\lambda)}_{\perp\mu}p_\nu-e^{(\lambda)}_{\perp\nu}p_\mu]
  \widetilde S(v,pz).}\hspace*{1.6cm}\label{eq:2.21}
\end{eqnarray}
Of these seven amplitudes, ${\cal T}$ is of twist-3 and the other six
of twist-4; higher twist terms are suppressed. 
In the above equations, we use
\begin{equation}
   {\cal T}(v,pz) =\int {\cal D}\underline{\alpha} 
e^{-ipz(\alpha_2-\alpha_1+v\alpha_3)}{\cal T}(\underline{\alpha}),
\end{equation}
etc., and $\underline{\alpha}$ is the set of three momentum fractions
$\underline{\alpha}=\{\alpha_1,\alpha_2,\alpha_3\}$.
 The integration measure is defined as 
\begin{equation}
 \int {\cal D}\underline{\alpha} \equiv \int_0^1 d\alpha_1
  \int_0^1 d\alpha_2\int_0^1 d\alpha_3 \,\delta\left(1-\sum \alpha_i\right).
\label{eq:measure}
\end{equation}
As for chiral-even DAs, to order $O(\delta^2)$ only the twist-3 DAs
contribute, which we define as
\begin{eqnarray}
\langle 0|\bar q_2(z) g\widetilde G_{\mu\nu}(vz)\gamma_\alpha\gamma_5 
  q_1(-z)|V(p,\lambda)\rangle & = &
  f_V m_V p_\alpha[p_\nu e^{(\lambda)}_{\perp\mu}
 -p_\mu e^{(\lambda)}_{\perp\nu}]{\cal A}(v,pz) +
  O(m_V^3),\hspace*{0.6cm}
\label{eq:even1}\\ 
\langle 0|\bar q_2(z) g G_{\mu\nu}(vz)i\gamma_\alpha 
  q_1(-z)|V(p)\rangle &=&
  f_V m_V p_\alpha[p_\nu e^{(\lambda)}_{\perp\mu} 
  - p_\mu e^{(\lambda)}_{\perp\nu}]{\cal V}(v,pz) +
  O(m_V^3).\hspace*{0.6cm}
\label{eq:even2}
\end{eqnarray}

At first glance, the sheer number of different DAs, 2 of twist-2, 7 of
twist-3 and 9 of twist-4, seems to preclude any predictivity of the
LCSRs. Appearances are deceiving, though: not all these DAs are
independent of each other, and one can disentangle their mutual
interdependencies using the {\em QCD equations of motion}, which results in
integral relations between different DAs, e.g.\ the chiral-odd DAs 
$\phi_\parallel$, $g_\perp^{(a,v)}$, $g_3$ etc.\ We shall see 
examples of such relations below. The other important organising
principle for DAs is {\em conformal expansion}, i.e.\ a partial wave 
expansion of DAs in terms of contributions of increasing
conformal spin. Conformal expansion relies on the fact that massless
QCD displays conformal symmetry\footnote{See Ref.~\cite{VBreview} for
 a review on the use of conformal symmetry in QCD.} 
which allows one to organise the DAs in
terms of irreducible representations of the corresponding symmetry
group SL$(2,{\mathbb R})$. The coefficients of these
different partial waves renormalize multiplicatively to LO in QCD, but
mix at NLO, the reason being that the symmetry is anomalous.

As mentioned above, the plethora of vector meson DAs is not mutually
independent, but related by the QCD equations of motion. These
relations are discussed at length in Ref.~\cite{wavefunctions}, whose
formulas we adapt to the present case. The chiral-even 
twist-3 DAs are of order
$\delta$, so we keep the full dependence on terms induced by
$\phi_\parallel$, but use conformal expansion for the admixture of
3-particle DAs:
\begin{eqnarray}
\left(1-\delta_+\right) g_\perp^{(a)} & = & \bar u
\int_0^u\,dv\,\frac{\Psi(v)}{\bar v} + u
\int_u^1\,dv\,\frac{\Psi(v)}{v}\nonumber\\
&&{}+ 10\zeta_3 \left(1 - \frac{3}{16}\,\omega_3^A +
\frac{9}{16}\,\omega_3^V \right) \left\{ 5
(2u-1)^2-1\right\},\\
g_\perp^{(v)} & = & \frac{1}{4}\left[
\int_0^u\,dv\,\frac{\Psi(v)}{\bar v} + 
\int_u^1\,dv\,\frac{\Psi(v)}{v}\right]+
5\zeta_3 \left\{ 3 (2u-1)^2-1\right\}\nonumber\\
&&{} + \frac{15}{64}\,\zeta_3
\left( 3 \omega_3^V - \omega_3^A\right) \left( 3 - 30 (2u-1)^2 + 35
(2u-1)^4 \right),
\end{eqnarray}
with $\Psi(u) = 2 \phi_\parallel(u) + \delta_+(2u-1)\phi'_\perp(u)
+\delta_- \phi'_\perp(u)$, $\delta_{\pm} = (f_V^T/f_V)\, (m_{q_2}\pm
m_{q_1})/m_V$. 
The dimensionless coupling $\zeta_3$ is
defined by the (local) matrix element
\begin{eqnarray}
\langle0|\bar q_2 g\tilde G_{\mu\nu}\gamma_\alpha
 \gamma_5 q_1|V(P,\lambda)\rangle & = & 
f_V m_V \zeta_{3}
\Bigg[
e^{(\lambda)}_\mu\Big(P_\alpha P_\nu-\frac{1}{3}m^2_V \,g_{\alpha\nu}\Big)
-e^{(\lambda)}_\nu\Big(P_\alpha P_\mu-\frac{1}{3}m^2_V \,g_{\alpha\mu}\Big)
\Bigg]\nonumber\\
& & {}+ \frac{1}{3}f_V m_V^3 \zeta_{4}
\Bigg[e^{(\lambda)}_\mu g_{\alpha\nu}- e^{(\lambda)}_\nu g_{\alpha\mu}\Bigg],
\label{def:zeta34}
\end{eqnarray}
where $\zeta_4$ is a matrix-element of twist-4. $\omega_3^{A,V,T}$ are
matrix elements of quark-quark-gluon operators involving derivatives
and defined in the second reference of \cite{wavefunctions}.

The chiral-odd twist-3 DAs, on the other hand, are $O(\delta^2)$, so
we model them in conformal expansion truncated after the first
non-leading order:
\begin{eqnarray}
h_\parallel^{(s)}(u) & = & 6u\bar u \left[ 1 + \left( \frac{1}{4}a_2^\perp +
\frac{5}{8}\,\zeta_{3}\omega_3^T \right) (5(2u-1)^2-1)\right],\\
h_\parallel^{(t)}(u) &= & 3(2u-1)^2+ 
 \frac{3}{2} a_2^\perp\, (2u-1)^2 \,(5(2u-1)^2-3)\nonumber\\
&&{}+\frac{15}{16}\zeta_{3}\omega_3^T(3-30(2u-1)^2+35(2u-1)^4).
\end{eqnarray}
As mentioned above, we drop contributions in the odd
Gegenbauer-moment $a_1^\perp$, as not
all $m_V^2 a_1^\perp$ terms are known.

As for the 3-particle twist-3 DAs, we have, quoting from 
Ref.~\cite{wavefunctions}:
\begin{eqnarray}
{\cal V} (\underline{\alpha}) &=& 
540\, \zeta_3 \omega^V_3 (\alpha_1-\alpha_2)\alpha_1 \alpha_2\alpha_3^2,
\label{modelV}\\
{\cal A} (\underline{\alpha}) &=& 
360\,\zeta_3 \alpha_1 \alpha_2 \alpha_3^2 
\left[ 1+ \omega^A_{3}\,\frac{1}{2}\,(7\alpha_3-3)\right],\label{modelA}\\
{\cal T}(\underline{\alpha}) &=& 540 \,\zeta_3\, \omega_3^T
(\alpha_1-\alpha_2) \alpha_1 \alpha_2 \alpha_3^2.\label{modelT}
 \end{eqnarray}
These expressions are valid to NLO in the conformal expansion.

\begin{table}
\renewcommand{\arraystretch}{1.4}
\addtolength{\arraycolsep}{3pt}
$$
\begin{array}{|c|ccc|}\hline
\mu & \zeta_3 & \omega_3^A & \omega_3^V\\ \hline
1\,{\rm GeV}\phantom{.2} & 0.032\pm 0.010 &
-2.1\pm1.0 & 3.8\pm1.8 \\ 
2.2\,{\rm GeV} & 0.018\pm 0.006 & -1.7\pm 0.9 & 3.6\pm 1.7 \\\hline
\end{array}
$$
\caption[]{3-particle parameters of chiral-even distribution
  amplitudes.}\label{tab:A} 
$$
\begin{array}{|c|ccccc|}\hline
\mu & \omega_3^T &  \zeta_4^T & \tilde{\zeta_4^T} & 
\langle\!\langle Q^{(1)}\rangle\!\rangle
 & \langle\!\langle Q^{(3)}\rangle\!\rangle \\ \hline
1\,{\rm GeV}\phantom{.2} & 7.0 \pm 7.0 & 0.10\pm 0.05 &
-0.10\pm 0.05 & -0.15\pm 0.15 & 0  \\ 
2.2\,{\rm GeV} & 7.2\pm 7.2 & 0.06\pm 0.03 & -0.06\pm 0.03 & -0.07\pm
0.07& 0 \\\hline
\end{array}
$$
\caption[]{3-particle parameters of
           chiral-odd distribution amplitudes. Terms in $a_2^\perp$
           are treated as described in Sec.~\ref{sec:4.3}.}\label{tab:B}
\renewcommand{\arraystretch}{1}
\addtolength{\arraycolsep}{-3pt}
\end{table}

The chiral-even 2-particle DAs of twist-4, $g_3$ and $\mathbb A_\parallel$ in
Eq.~(\ref{eq:OPEvector}),  are $O(\delta^3)$, so we drop them. For the
chiral-odd twist-4 DAs $h_3$ and ${\mathbb A}_\perp$  we use NLO conformal
expansion (with $a_1^\perp\to 0$):
\begin{eqnarray}
h_3(u) & = & 1 + \left\{\frac{3}{7}\,a_2^\perp -1 - 10 (\zeta_4^T +
  \wt{\zeta}_4^T ) \right\}\! C_2^{1/2}(2u-1)+ \left\{
   -\frac{3}{7}\, a_2^\perp - \frac{15}{8}\, \zeta_3
  \omega_3^T\right\}\! C_4^{1/2}(2u-1),\nonumber\\[-0.5cm]\label{eq:h3exp}
\end{eqnarray}
\begin{eqnarray}
{\mathbb A}_\perp(u) & = & 30 u^2 \bar u^2 \left\{ \frac{2}{5} \left( 1 +
  \frac{2}{7}\, a_2^\perp + \frac{10}{3}\, \zeta_4^T - \frac{20}{3}\,
  \wt{\zeta}_4^T \right) + \left( \frac{3}{35}\, a_2^\perp +
  \frac{1}{40}\, \zeta_3 \omega_3^T \right) C_2^{5/2}(2u-1)
  \right\}\nonumber\\
 & & {} - \left( \frac{18}{11}\, a_2^\perp   - \frac{3}{2}\, \zeta_3
  \omega_3^T + \frac{126}{55}\,
 \langle\!\langle Q^{(1)}\rangle\!\rangle + \frac{70}{11}\, 
\langle\!\langle Q^{(3)}\rangle\!\rangle
   \right)\nonumber\\
 & & \times \left( u\bar u (2+13 u\bar u) + 2u^3 (10-15u+6u^2) \ln u +
  2\bar u^3 (10-15\bar u + 6\bar u^2) \ln \bar u\right).\hspace*{0.9cm}
\label{eq:ATexp}
\end{eqnarray}
The formulas for chiral-odd 3-particle DAs of twist-4 are rather
lengthy and we refrain from reproducing them here. They can be
found in the second reference of \cite{wavefunctions}. 

The numerical values of  3-particle matrix-elements are given in
Tabs.~\ref{tab:A} and \ref{tab:B}, for the scales 1~GeV and 
$\sqrt{m_B^2-m_b^2} = 2.2\,$GeV. The corresponding one-loop anomalous
dimensions are also given in \cite{wavefunctions}. The numerical
values for the decays constants $f_V^{(T)}$ are collected in Tab.~\ref{tab:fV}.

\section{3-Particle Contributions to the LCSRs}\label{app:B}

In this paper we include contributions of 3-particle
DAs to the correlation function (\ref{eq:corr}) at tree level. This appendix contains
  explicit formulas for these contributions.

The 3-particle DAs of twist-3 have been defined in
  App.~\ref{app:A}; the definitions for twist-4 DAs can be found in Ref.~\cite{wavefunctions}.
Their contributions to the correlation functions are most easily
  calculated in the external field method proposed in Ref.~\cite{NSVZ}.
The light-cone $b$ quark propagator in an external field reads, 
in the Fock-Schwinger gauge $x_\mu A^\mu(x)=0$:
\begin{equation}
\matel{0}{Tb(x)\bar{b}(0)}{0}_A = iS_b^{(0)}(x)+iS_b^{(2)}(x,0),
\end{equation}
with
\begin{equation}
S_b^{(2)}(x,0)= -\int \frac{d^4k}{(2\pi)^4} e^{-ik\cdot x} \int_0^1 dv
\frac{1}{2}\big( \bar{v}S_2(k,m)\, \sigma_{\mu\nu} G^{\mu\nu}(vx) + v \sigma_{\mu\nu} 
G^{\mu\nu}(vx)\, S_2(k,m) \big),
\end{equation}
where $S_n(k,m) = (\s{k}+m)/\Delta^n(k)$ with $\Delta(k)\equiv
1/(k^2-m^2)$.\footnote{Note
that $S^{(2)}(x,0) \neq S^{(2)}(0,-x)$ as the Fock-Schwinger gauge
breaks translational invariance.} 
This expression  is equivalent to Eq.~(2.25) in Ref.~\cite{BB98}.
The  decomposition \eqref{20A} selects the chiral odd  DAs \eqref{eq:T3}, 
\eqref{eq:2.21} and
the chiral even  DAs \eqref{eq:even1}, \eqref{eq:even2}. 
Terms in $e^*_\al x_\beta/px$ are treated by partial integration;
we have checked that all boundary terms vanish. 
Upon partial integration, we hence have
\begin{equation*}
\frac{e^*_\al x_\beta}{px} \int_\al e^{ix \cdot (k-l)}
f(\al_1,\al_3) 
S_2(k,m) \to
\int_\al e^{ix \cdot (k-l)} f(\tilde{\al}_1,\al_3) \Big[ 4
  S_3(k,m)e^*_\al 
k_\beta 
-e^*_\al \ga_\beta \Delta(k)^2 \Big] \quad ,
\end{equation*}
with $l= q+ (\al_1+v \al_3)p$, $f(\tilde{x},y) = \int^x da f(a,y)$
and 
$\int_\al =  
\int_0^1 d\al_3 \int_0^{1-\al_3} d\alpha_1$.

The contribution of 3-particle DAs to the
correlation function \eqref{eq:corr} then reads:
\begin{eqnarray*}
\frac{i}{4} f_V m_V & & \int_0^1 dv \int D\al\, \Delta(l)^2\,  
(pq)\big({\cal V}(\al)+{\cal A}(\al)\big) \,
2 v {\rm tr}[\Gamma \s{e}^* \s{p} \ga_5] + O(m_V^3)\\
+&&\frac{i}{4} f_V^T m_V^2 \left(\int_0^1 dv \int D\al\,
\Delta(l)^2\, S(\al)  
(\bar{v} {\rm tr}[\Gamma (\s{q}+m)  \s{e}^*\s{p}\ga_5]+v{\rm
  tr}[\Gamma 
\s{e}^*\s{p}(\s{q}+m) \ga_5])\right.
\nonumber \\
-& &  \int_0^1 dv \int D\al\, \Delta(l)^2\, \tilde{S}(\al)  
(\bar{v}{\rm tr}[\Gamma (\s{q}+m)  \s{e}^*\s{p}\ga_5]+v{\rm
  tr}[\Gamma 
\s{e}^*\s{p}(-\s{q}+m)\ga_5])\nonumber\\
+& &  \int_0^1 dv \int D\al\, \Delta(l)^2\,
\bar{v}{\rm tr}[\Gamma (\s{q}+m)  \s{e}^*\s{p}\ga_5]
  T^{(4)}_3(\al_1,\al_3) 
\nonumber \\
+& &  \int_0^1 dv \int D\al\, \Delta(l)^3\,
4v (pq){\rm tr}[\Gamma (-\s{q}+m)  \s{e}^*\s{p}\ga_5] T^{(4)}_3(
\tilde{\al_1},\al_3) \nonumber \\
-& & \int_0^1 dv \int D\al\, \Delta(l)^2\,
\bar{v}{\rm tr}[\Gamma (\s{q}+m)  \s{e}^*\s{p}\ga_5]
  T^{(4)}_4(\al_1,\al_3)
 \nonumber 
\end{eqnarray*}
\begin{eqnarray*}
-& &  \int_0^1 dv \int D\al\, \Delta(l)^3\,
4v (pq){\rm tr}[\Gamma \s{e}^*\s{p}  (-\s{q}+m)  \ga_5] T^{(4)}_4(
\tilde{\al_1},\al_3) \nonumber \\
+& &  \int_0^1 dv \int D\al\, 
\Big[ (16\bar{v}\{p(e^*q)\} + 4v\{e^*pq\}-4vm\{e^*p\})
  \,(pq)
\Delta(l)^3\,T^{(4)}_1(\tilde{\al_1},\al_3) 
\nonumber
\end{eqnarray*}
\begin{eqnarray*}
+& &
  ((\bar{v}-v)\{qe^*p\}-4v\{e^*(pq)\}+v\{e^*pq\}+m(\bar{v}+2v)\{
e^*p\})\,\Delta(l)^2\,T^{(4)}_1
(\al_1,\al_3) \Big] 
\nonumber  \\
+& &   \int_0^1 dv \int D\al\, 
\Big[ (-16\bar{v}\{p(e^*q)\} - 4v\{qe^*p\}-4vm\{e^*p\})
  \,(pq)
\Delta(l)^3\,T^{(4)}_2(\tilde{\al_1},\al_3) 
\nonumber \\
+& &\left.\left(
  ((\bar{v}+v)\{qe^*p\}-4v\{e^*(pq)\}+v\{e^*pq\}+m(\bar{v}+2v)\{
e^*p\})\,\Delta(l)^2\,T^{(4)}_2
(\al_1,\al_3) \Big]\right.\right). \nonumber
\end{eqnarray*}
In the above formula, we use $\{abc\}={\rm tr}[\Gamma \s{a}\s{b}\s{c} \ga_5]$ and 
$\{a(bc)\} = b \cdot c\;{\rm tr}[\Gamma \s{a} \ga_5]$.

\end{document}